\documentclass[aps,prd,a4paper,10ot,amsmath,amssymb,nofootinbib,%
showpacs,showkeys,notitlepage,superscriptaddress,twocolumn]{revtex4-1}
\usepackage{bm}
\usepackage{slashbox}
\usepackage[FIGTOPCAP]{subfigure}

\usepackage[dvipdfm]{graphicx,color}
\usepackage{mathrsfs}

\newcommand{\disp}{\displaystyle}
\newcommand{\ie}{{i.e.}}
\newcommand{\beq}{\begin{equation}}
\newcommand{\eeq}{\end{equation}}
\newcommand{\ba}{\begin{array}}
\newcommand{\ea}{\end{array}}
\DeclareMathOperator{\re}{Re}
\DeclareMathOperator{\im}{Im}
\DeclareMathOperator{\sgn}{sgn}

\begin{document}
\title{Crystalline chiral condensates off the tricritical point in a
generalized Ginzburg-Landau approach}
\author{Hiroaki Abuki}
\email{h.abuki@rs.tus.ac.jp}
\affiliation{Department of Physics, Tokyo University of Science,
Kagurazaka 1-3, Shinjuku, Tokyo 162-8601, Japan}
\author{Daisuke Ishibashi}
\email{d.ishibashi.1@gmail.com}
\affiliation{Bureau of Waterworks, Tokyo Metropolitan Government,
Hongo 2-7-1, Bunkyo-ku, Tokyo 163-8001, Japan}
\author{Katsuhiko Suzuki}
\email{katsu\_s@rs.kagu.tus.ac.jp}
\affiliation{Department of Physics, Tokyo University of Science,
Kagurazaka 1-3, Shinjuku, Tokyo 162-8601, Japan}

\begin{abstract}
We present an extensive study on inhomogeneous chiral condensates in
 QCD at finite density in the chiral limit using a generalized
 Ginzburg-Landau (GL) approach.
Performing analyses on higher harmonics of one-dimensionally (1D)
 modulated condensates, we numerically confirm the previous claim
 that the solitonic chiral condensate characterized by Jacobi's
 elliptic function is the most favorable structure in 1D modulations.
We then investigate the possibility of realization of several
 multidimensional modulations within the same framework.
We also study the phase structure far away from the tricritical point by
 extending the GL functional expanded up to the eighth order in the order
 parameter and its spatial derivative.
On the same basis, we explore a new regime in the extended GL parameter
 space and find that the Lifshitz point is the point where five critical
 lines meet at once.
In particular, the existence of an intriguing triple point is
 demonstrated, and its trajectory consists of one of those critical
 lines.
\end{abstract}

\maketitle
\section{Introduction}
How matter behaves as a whole under extreme conditions, such as high
temperature and/or density is one of the central questions in hadron physics
that, in principle, should be answered by quantum chromodynamics (QCD).
At high temperature QCD unambiguously predicts a new,
extreme form of matter called the quark-gluon plasma, 
and its properties are now under extensive investigation from both
theoretical and experimental sides, the latter facilitated by heavy ion
colliding facilities such as RHIC and the LHC. 
On the theoretical side, the most powerful tool to investigate this
region is the lattice QCD simulations which makes possible {\it ab
initio} calculation of thermodynamic quantities at high temperature.

On the other hand, at finite density lattice simulations still remain
less predictive because of a serious problem due to the complex fermion
determinant.
Our knowledge of the forms of matter at finite density is limited except
for ultimately a high density regime where 
we know, from perturbative QCD computations, matter forms a
color superconductor of the color-flavor locked type.
Even though the form of matter at accessible density is still obscure,
we might expect to have two kinds of phase transition (or
crossover) when ordinary nuclei is compressed by an external
gravitational pressure: One is the deconfinement \cite{Collins:1974ky} 
and the other is the chiral restoration \cite{Hatsuda:1994pi} which is
the main topic of this paper.

QCD vacuum is considered as a sea of the chiral condensate
\cite{Nambu:1961fr}, namely, $\bar{q}q$ condensate which may arise due
to a strong correlation between a quark and antiquark. 
The assumption has proven to be useful for interpretation of 
hadron spectroscopy as well as hadron scatterings at low energy, 
yet it is only recently that its formation in vacuum was directly
checked by lattice simulations of QCD \cite{Fukaya:2010na}. 
When the baryon chemical potential is introduced, it breaks the
charge conjugation symmetry, bringing an imbalance, \ie, a net excess
of quarks over antiquarks.
Thus it is possible for the chiral condensate to be broken by such
imbalance. 
In the model calculation this actually takes place through the 
suppression of $\bar{q}q$ formation by the Pauli blocking
\cite{Hatsuda:1994pi,Klevansky:1992qe}. 

How the entire chiral restoration takes place is, however, still under
the veil. 
Most of the Nambu--Jona-Lasinio (NJL)-type models predict the first-order
phase transition. 
But it is possible that this transition proceeds hierarchically via
several intermediate steps, such as meson condensates.
In particular, possibilities of inhomogeneous chiral condensate
[chiral density waves (CDW)] have been attracting a lot of interest
recently
\cite{%
Deryagin:1992rw,Shuster:1999tn,%
Sadzikowski:2000ap,Nakano:2004cd,%
Nickel:2009ke,*Nickel:2009wj}. 
So far proposed CDWs can be classified into two major groups: 
one is the Flude-Ferrell (FF)-type \cite{FludeFerrell} and the other is
the Larkin-Ovchinnikov (LO) type \cite{larkin:1964zz,*larkin:1965sp}.
CDWs in the former group are characterized by some anisotropic pairing
in momentum space, while those in the latter group are described by
a characteristic modulation in real space.

CDWs can be regarded as femtoscale phase separations
as a consequence of the competition between the chiral condensate
$\sigma=\langle\bar{q}q\rangle$ and quark chemical potential $\mu$;
the latter drives a stress that promotes the population of quarks,
$n=\langle q^\dagger q\rangle$, that is, an imbalance in the sea of
quark-antiquark condensate.
In the case of the FF-type CDWs, quarks as impurities are accumulated 
in some favorable directions on the Fermi surface where the pairing
is absent, while in the LO-type CDWs the condensate modulates in real
space, taking nodes where quarks are populated without any friction
\cite{Carignano:2010ac}. 
In a nutshell, it can be said that the CDW is a {\it halfway} state
produced by a compromise between two conflicting effects: the $\bar{q}q$
condensate and the quark occupation $q^\dagger q$.

Essentially the same picture applies to conventional FFLO
superconductors in condensed matter physics
\cite{FludeFerrell,larkin:1964zz,*larkin:1965sp}. 
It was recently reported that a kind of such
exotic superconductors was observed experimentally in 2D films of a
heavy fermion compound, CeCoIn${}_5$ \cite{MatsudaFFLO}.
In these systems, an external magnetic field $h$ drives the
Pauli's paramagnetism via yielding a mismatch in the Fermi momenta of two
species, in this case, spin up and down. 
This paramagnetic stress competes with the spin singlet pairing, and as
a consequence, the FFLO state may appear as a halfway state between a
complete paramagnetic state and a uniform pairing state;
when $h$ exceeds some critical value, the spin density $n_3({\bf
x})=\langle f^\dagger\tau_3 f\rangle=n_{\uparrow}({\bf
x})-n_{\downarrow}({\bf x})$
starts to accumulate near the points where spin singlet pair density
$\Delta({\bf x})=\langle
f^T(i\tau_2)f\rangle=\langle\uparrow\downarrow-\downarrow\uparrow\rangle$
takes nodes in the real (or momentum) space.

Possible formation of inhomogeneous condensates has been discussed in
various contexts of high energy physics. These include neutral pion
condensates \cite{Kunihiro:1979yp,*Kunihiro:1985yx}, charged pion
condensates \cite{Ebert:2011rg}, color superconductors
\cite{Alford:2000ze,Bowers:2001ip,Leibovich:2001xr,Casalbuoni:2003wh,%
*Casalbuoni:2004wm,*Casalbuoni:2005zp,Nickel:2008ng},
chiral magnetic spiral in a strong magnetic field \cite{Basar:2010zd},
crystal phases in $(1+1)$-dimensional QCD in the large $N_c$ limit
\cite{Bringoltz:2009ym}, and spiral phases in the quarkyonic phase
\cite{Kojo:2009ha,Kojo:2010fe,Kojo:2011cn}.
Inhomogeneous chiral condensates have also been studied extensively
in the Gross-Neveau (GN) model with a discrete chiral symmetry
\cite{Schon:2000qy,Schnetz:2004vr,Schnetz:2005ih,%
Boehmer:2007ea,Basar:2008im,Basar:2008ki}
and also in the corresponding model with a continuous chiral
symmetry, \ie the $(1+1)$-dimensional NJL model in the large $N$ limit
\cite{Basar:2009fg}.

Even if the crystal structure is restricted to 1D modulations, its
impact in the QCD phase diagram is drastic as demonstrated in
\cite{Nickel:2009ke}; it was shown that there is a domain in the phase
diagram where the chiral condensate develops a crystallization in a
solitonic form characterized by the elliptic function. 
Accordingly, the tricritical point turns into the Lifshitz point where
three phases, a symmetric (Wigner) phase, an inhomogeneous phase, and a
homogeneous symmetry-broken phase, meet at once. 
The analyses were based on the Ginzburg-Landau (GL) functional expanded
in the chiral order parameter and its spatial derivatives up to
sixth order, which offers a minimal model-independent description of
the QCD tricritical point in the chiral limit \cite{Asakawa:1989bq}.
Despite the development in understanding the 1D modulated chiral
condensate, only a few works have been devoted to exploring
multidimensional modulations \cite{Kojo:2011cn,Carignano:2011gr}. 

The aim of the present paper is twofold: first to seek the most
favorable crystallization pattern and, second,
to explore the phase structure away from the tricritical point.
In connection with the first point, we note that even in the GL
framework it is not so evident that the solitonic chiral crystal
characterized by Jacobi's elliptic function gives the absolute
ground state. What is shown is that it certainly constitutes a
sufficient, particular solution at sixth order of the GL functional. 
We thus examine other crystal structures as candidates of the ground
state, including the possibility of multidimensional crystal structures. 
For a 1D structure, we search for the most favorable structure taking
the most general {\it Ansatz}, the condensate expanded in the harmonic
series \cite{Combescot:2005tt,Nickel:2008ng}.
As for a higher dimensional structure, we only examine several specific
{\it Ans\"atze}.

In order to accomplish the second purpose, we extend the existing GL
functional up to eighth order in the condensate and its spatial
derivative. 
This is because the GL functional at sixth order provides us
with a minimal description of the critical point, where it follows from
a simple scaling rule that the phase structure stays the same
even if the region close to the critical point is zoomed out. 
Moreover there is one more positive reasoning for this extension: 
It was shown in the context of a superconductor under an external magnetic
field that such higher order terms should be taken care to correctly
describe phases off the tricritical point \cite{Matsuo:1998pd}. 
Starting from the BCS model and making use of a quasiclassical
(Eilenberger) approximation to the Bogoliubov--de Genne equation,
they numerically found that another type of FFLO phase named FFLO-II
shows up in the phase diagram; it replaces a large part of the
inhomogeneous phase, which brings a new critical end point located off
the tricritical point.
In order to understand their results within the GL framework, one needs
to work at least at eighth order as pointed out in the paper.

The paper is organized as follows. Section~\ref{sec:review} is mostly
devoted to a review on the GL treatment of inhomogeneous chiral
condensates made in \cite{Nickel:2008ng}.
We make physics behind a formation of inhomogeneous condensate
transparent. In Sec.~\ref{sec:multi}, we examine several
{\it Ans\"atze} for multidimensional crystal structures, as well as the most
general {\it Ansatz} for 1D modulation.
We finally address in Sec.~\ref{sec:offcri} the question of how the phase
structure expands at large distance from the Lifshitz point by extending
the previous GL functional up to eighth order.
We summarize the present paper in Sec.~\ref{sec:summary}.

\section{Generalized Ginzburg-Landau approach on inhomogeneous 
                phases}\label{sec:review}
Nickel studied the possibility of inhomogeneous chiral phases in QCD using
a generalized GL approach \cite{Nickel:2009ke}.
Here we review the framework.
In order to describe the phase structure close to the tricritical point, 
we need to retain up to sixth order in the
chiral order parameter in the GL Lagrangian. 
Following \cite{Nickel:2009ke}, we extend the GL Lagrangian such
that the theory allows an inhomogeneous phase.
\subsection{Generalized Ginzburg-Landau functional}
To incorporate the energy gain/loss with respect to space variations of
the order parameter, we employ the gradient expansion.
A generalized GL functional in the chiral limit up to sixth order in
the chiral order parameter and its derivative is given by
\beq
\ba{rcl}
  \omega(M({\textbf x}))&=&\frac{\alpha_2}{2}M({\textbf x})^2%
 +\frac{\alpha_4}{4}M({\textbf x})^4+\frac{\alpha_{4b}}{4}%
 (\nabla M({\textbf x}))^2\\[2ex]
  & &+\frac{\alpha_6}{6}M^6+\frac{\alpha_{6b}}{6}%
 M^2(\nabla M)^2%
 +\frac{\alpha_{6c}}{6}(\Delta M)^2.
\ea
\eeq
Nickel derived, starting from an NJL model, relations among parameters
$\{\alpha_n\}$ up to total derivatives: $\alpha_{4b}=\alpha_4$,
$\alpha_{6b}=5\alpha_6$ and $\alpha_{6c}=\alpha_6/2$. 
Hereafter, we treat these three parameters
$\{\alpha_2,\alpha_4,\alpha_6\}$ as independent GL couplings. 
In the NJL model, only quark loops contribute to the GL couplings, which may
provide a good description at high density regime.
Then the generalized GL functional to this order becomes
\beq
\ba{rcl}
 \omega(M({\textbf x}))&=&\frac{\alpha_2}{2}M({\textbf x})^2%
 +\frac{\alpha_4}{4}(M({\textbf x})^4+(\nabla M)^2)\\[2ex]
 &&+\frac{\alpha_6}{6}\left(M({\textbf x})^6+5M^2(\nabla
 M)^2+\frac{1}{2}(\Delta M)^2\right).\\[1ex]
\ea
\label{eq:GLlag}
\eeq
We can then analyze the phase structure of quark matter in the GL
parameter space spanned by $\{\alpha_2,\alpha_4,\alpha_6\}$, which can
in principle be mapped onto the phase diagram in the $(\mu,T)$ space in
real QCD.

\subsection{Dimensional and scaling analysis}
Parameter $\alpha_6$ should always be positive for the thermodynamic
stability and it has dimension $\Lambda^{-2}$ with $\Lambda$ being
some energy scale.
Then we use $\alpha_6$ to specify the unit of energy scale.
Moreover we can use $|\alpha_4|$ to adjust the magnitudes of the 
fourth- and sixth-order coefficients. 
To be precise, let us introduce dimensionless variables
$\{\eta_2,\omega,m,\tilde{{\bf x}},\tilde{\nabla}\}$ as follows:
\beq
\ba{rcl}
  \alpha_2&=&\eta_2\,\left[\frac{\alpha_4^2}{\alpha_6}\right],\\[2ex]
  \omega&=&\tilde{\omega}\,\left[\frac{|\alpha_4|^3}{\alpha_6^2}\right],\quad
  M=m\,\left[\frac{|\alpha_4|^{1/2}}{\alpha_6^{1/2}}\right],\\[2ex]
  {\bf x}&=&\tilde{\bf
  x}\,\left[\frac{\alpha_6^{1/2}}{|\alpha_4|^{1/2}}\right],%
  \quad \tilde{\nabla}=\nabla_{\tilde{\bf x}}.\\[1ex]
\ea
\label{eq:scaling}
\eeq
Then, in these dimensionless bases, the GL functional takes the form
\beq
\ba{rcl}
   \tilde{\omega}&=&%
  \frac{\eta_2}{2}m^2%
  +\frac{\sgn(\alpha_4)}{4}%
  \left({m}^4+(\tilde\nabla m)^2%
  \right)\\[2ex]
  &&+\frac{1}{6}\left(m^6%
  +5m^2(\tilde\nabla m)^2%
  +\frac{1}{2}(\tilde\Delta m)^2\right).
\ea
\label{eq:omega}
\eeq
Now we see that the problem has only one continuous parameter $\eta_2$.
What we have to work out now is to find the phase structure in
$\eta_2$ space both for positive and negative $\alpha_4$ cases, and 
then map them back onto the original GL parameter space
$\{\alpha_2,\alpha_4,\alpha_6\}$ according to the scaling relation
between $\alpha_2$ and $\eta_2$ in Eq.~(\ref{eq:scaling}).

In the following we use $\{\alpha_2$, $M$, ${\bf x}\}$
instead of $\{\alpha_2\alpha_6$, $M\sqrt{\alpha_6}$, ${\bf
x}/\sqrt{\alpha_6}\}$ to avoid notational complications. 
They should be understood as dimensionless ones measured in the proper
units, \ie, $[\alpha_6^{-1}]$, $[\alpha_6^{-1/2}]$ and $[\alpha_6^{1/2}]$.

\subsection{Condition for soliton/condensate formation}
For $\alpha_4>0$ the phase structure is simple. One has a uniform chiral
symmetry-broken phase ($M\ne0$, $\chi$SB phase) for $\eta_2<0$ and
symmetry restored phase ($M=0$, Wigner phase) for $\eta_2>0$, separated
by a second-order phase transition. 
On the other hand, phases for $\alpha_4<0$ are rather rich
\cite{Nickel:2009ke}. 
When the condensate is restricted to be constant in space, there are
only two phases: the $\chi$SB phase at small $\eta_2$ and the Wigner
phase at large $\eta_2$, separated by a first-order phase transition at
$\eta_2=3/16\equiv \eta_2^{\mathrm{c}}$. 
Once the chiral condensate is allowed to have variations in space, the
first-order phase transition splits into two second-order phase
transitions, one at $\eta_2^{\mathrm{I}}=5/36<\eta_2^c$, the other at
$\eta_2^{\mathrm{II}}=3/8>\eta_2^c$, and in between an inhomogeneous
phase shows up \cite{Nickel:2009ke} as we will discuss below.

Let us start with a one-dimensional sinusoidal chiral density wave
\cite{Deryagin:1992rw,Shuster:1999tn} (the LO state) characterized by
\beq
  M_{\mathrm{LO}}(\textbf{x})=\sqrt{2}M_\mathrm{II}\sin(k_{\mathrm{II}} z).
\label{eq:sin}
\eeq
Averaging the GL free energy density over the Wigner-Seitz cell,
optimizing the wave vector $k_{\mathrm{II}}$ with respect to
$M_\mathrm{II}$, and finally  expanding the GL free energy in
$M_\mathrm{II}$, we obtain for $\alpha_4<0$, up to quartic order in
$M_\mathrm{II}$,
\beq
\ba{rcl}
\Omega&=&\langle\omega(M_{\mathrm{LO}}(\textbf{x}))%
\rangle_{\mathrm{WS}}\\[1ex]
&=&\left(\frac{\eta_2}{2}-\frac{3}{16}\right)M_\mathrm{II}^2%
 +\frac{1}{4}M_\mathrm{II}^4+{\mathcal O}(M_\mathrm{II}^6).
\ea
\label{eq:GLexforLO}
\eeq
From this, it is clear that the system undergoes a second-order
transition from the Wigner phase to the LO state when the parameter
$\eta_2$ crosses $\eta_2^{\mathrm{II}}=3/8$ from above.

\begin{figure}[tbp]
\centering
\subfigure[\hspace*{-3.0em}]{\includegraphics[scale=0.6,clip]%
{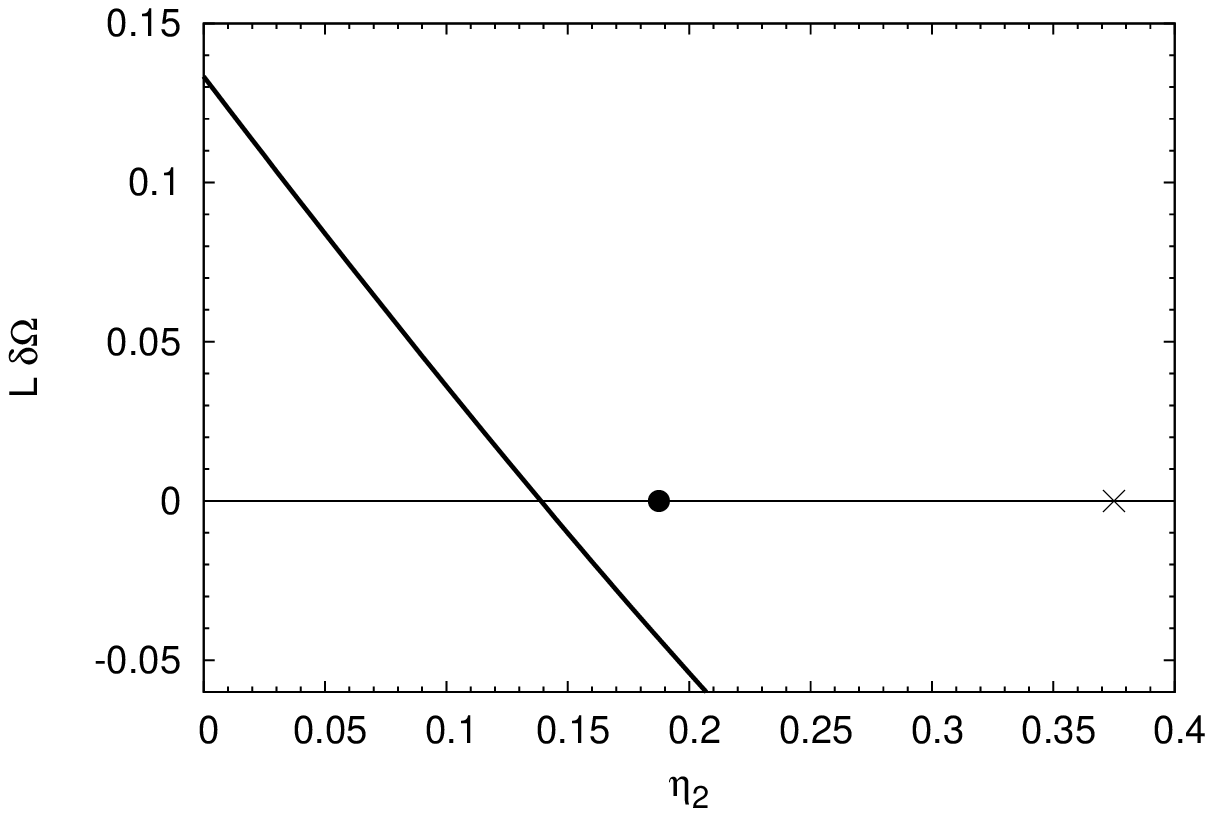}}
\hspace*{1.8em}
\subfigure[]{\includegraphics[scale=0.65,clip]%
{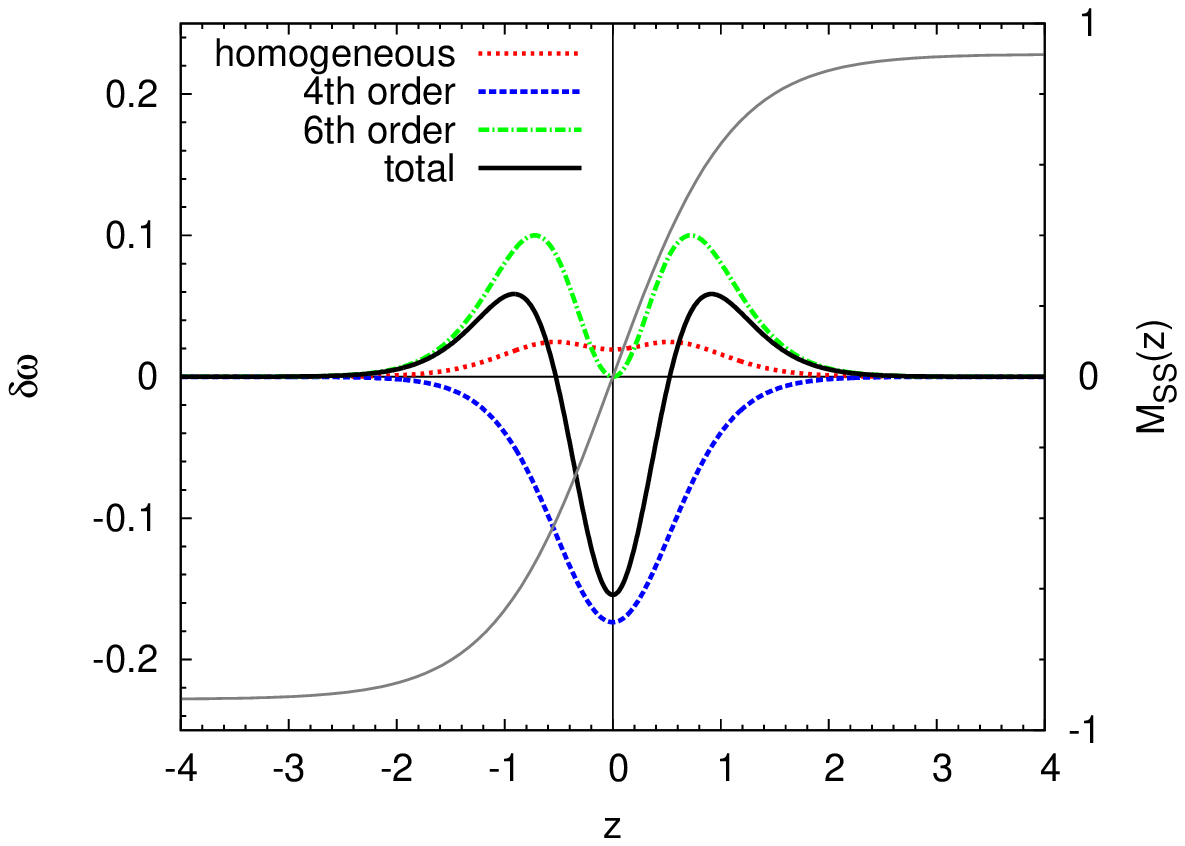}}
\caption{(color online). {\bf (a)}~Energy per unit area required for the creation of
 a single soliton as a function of $\eta_2$. 
 The solid circle and the cross placed on the $\eta_2$ axis represent
 points $\eta_2=\eta_2^{\mathrm{I}}$ and $\eta_2=\eta_2^{\mathrm{c}}$. 
 See text for details.
{\bf (b)}~A single soliton profile
 $M_{\mathrm{SS}}=M_{\mathrm{I}}\tanh(M_\mathrm{I}z)$ 
 with $M_\mathrm{I}=\sqrt{{5}/{6}}$ being the mass at
 $\eta_2=\eta_2^{\mathrm{I}}$ and the associated energy density profile. 
 Contributions from homogeneous parts, and those from derivative terms
 (at the fourth and sixth orders separately), are also shown.}
\label{fig:1}
\end{figure}

Now we move on to the region of the lower value of $\eta_2$, where the
chiral symmetry is broken by a homogeneous condensate.
Consider the situation where a single soliton (a kink in 1D
\cite{Schon:2000qy,Boehmer:2007ea,Basar:2008im},
actually a domain wall in 3D \cite{Nickel:2009ke,*Nickel:2009wj}) is
formed in a sea of homogeneous condensate. 
When $\eta_2$ is small, there is an energy cost for the creation of such
a kink in the homogeneous condensate.
However, this energy cost decreases with increasing $\eta_2$,
and at some critical point which we denote by $\eta_2^{\mathrm{I}}$ it
vanishes.
To see this, we first set an {\it Ansatz} for the domain wall profile
as
\beq
  M_{\mathrm{SS}}(z)=M_{\mathrm{I}}\tanh(k_{\mathrm{I}} z),
\eeq 
where $M_{\mathrm{I}}$ is the homogeneous condensate, being a function
of $\eta_2$.
The energy cost accumulated within the domain wall surface per unit
area can be defined by
\beq
\ba{rcl}
 \ell\Omega_\ell(M_{\mathrm{I}},k_{\mathrm{I}};\eta_2)&\equiv&%
 \int_{-\ell/2}^{\ell/2}dz(\omega(M_{\mathrm{SS}}({z}))%
 -\omega(M_\mathrm{I})).\\[1ex]
\ea
\eeq
where $\ell$ is the system length in the $z$ direction which will be
taken to be infinity.
Optimizing the value of $k_{\mathrm{I}}$ requiring that the energy cost
$F(\eta_2,k_{\mathrm{I}})=\lim_{\ell\to\infty}%
[\ell\delta\Omega_{\ell}(M_\mathrm{I},k_{\mathrm{I}};\eta_2)]$
takes minimum, we obtain the relation $k_{\mathrm{I}}=M_\mathrm{I}$. 
The function $f(\eta_2)\equiv F(\eta_2,M_{\mathrm{I}}(\eta_2))$
is depicted in Fig.~\ref{fig:1}(a), from which
we see that it is a monotonically decreasing function of $\eta_2$,
and at the critical point which we denote by $\eta_2^{\mathrm{I}}$ it
crosses zero. 
At this point, a soliton will be formed spontaneously without any energy
cost. It is easy to derive $\eta_2^{\mathrm{I}}=5/36$ where
$M_I=\sqrt{5/6}$, and in fact $f(\eta_2)$ is expanded in powers of
$\Delta\equiv(\eta_2-{5}/{36})$, as $f(\eta_2)=-\frac{5}{6}\Delta%
+\frac{9}{8}\Delta^2+{\mathcal O}(\Delta^3)$.
In the figure, we denote by a solid circle $\eta_2^\mathrm{c}=3/16$,
and by the cross $\eta_2^{\mathrm{II}}=2\eta_2^{\mathrm{c}}$ for comparison.
Obviously the domain wall formation is faster to take place than the
first-order phase transition to the Wigner phase when $\eta_2$ is
increased. We conclude that
some kind of inhomogeneous phase should develop in the window
$\eta_2^{\mathrm{I}}<\eta_2<\eta_2^{\mathrm{II}}$.

In order to elucidate on what drives the formation of a domain wall in the
spatially uniform $\chi$SB phase at the critical point $\eta_2^{\mathrm{I}}$, 
we show in Fig.~\ref{fig:1}(b) the energy density profile of a single
soliton by the bold line. 
The spatial profile of the soliton is also shown by the thin line. 
Several contributions to the energy density are shown separately:
The dotted (red) line shows the energy loss through the 
homogeneous parts in the GL functional,  
while the dashed (blue) and dot-dashed (green) lines
show the contribution from the derivative terms at the fourth and
sixth orders, respectively. We confirm that the derivative term at the
fourth order is responsible for the formation of a kink in the chiral
condensate.

\subsection{Solitonic  solution at sixth order of GL expansion}
\begin{figure}[tbp]
\centering
\subfigure[]{\includegraphics[scale=0.6]{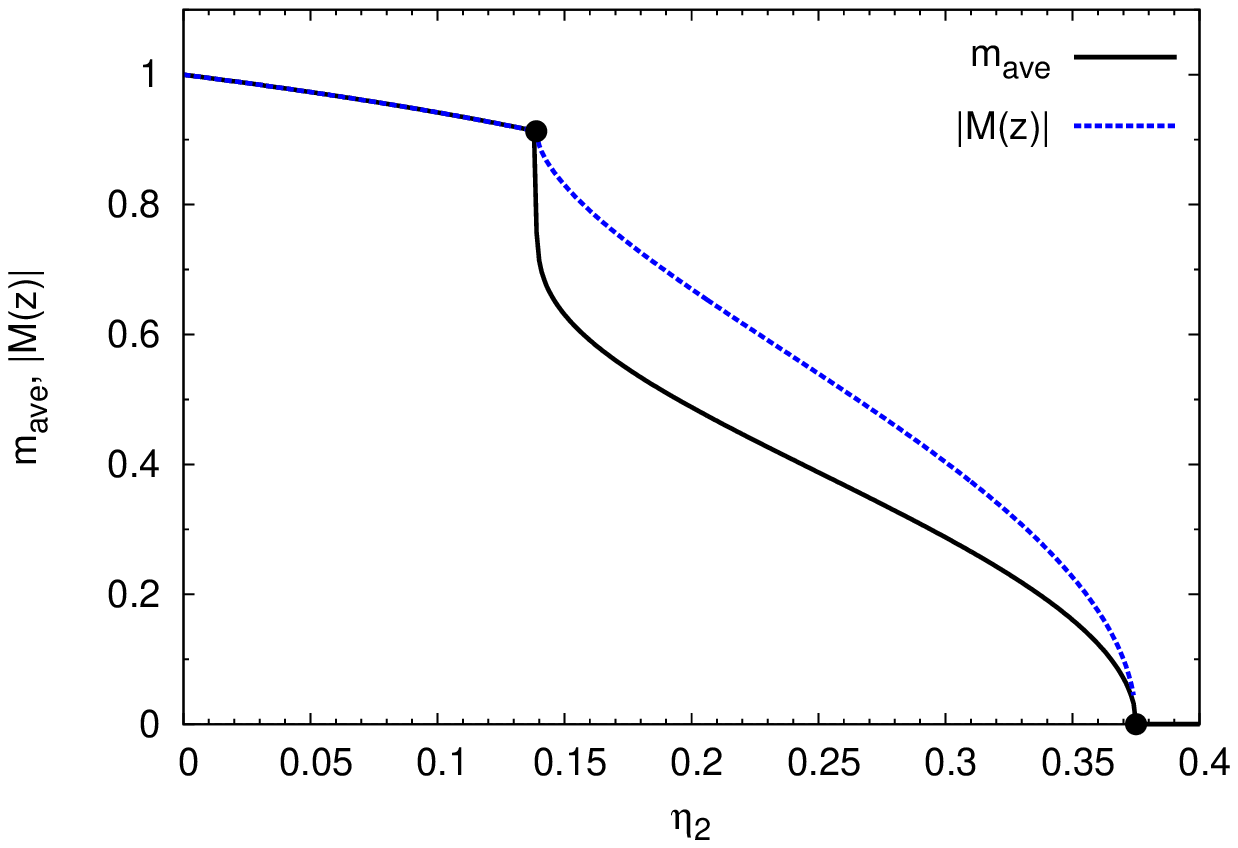}}
\subfigure[]{\includegraphics[scale=0.6]{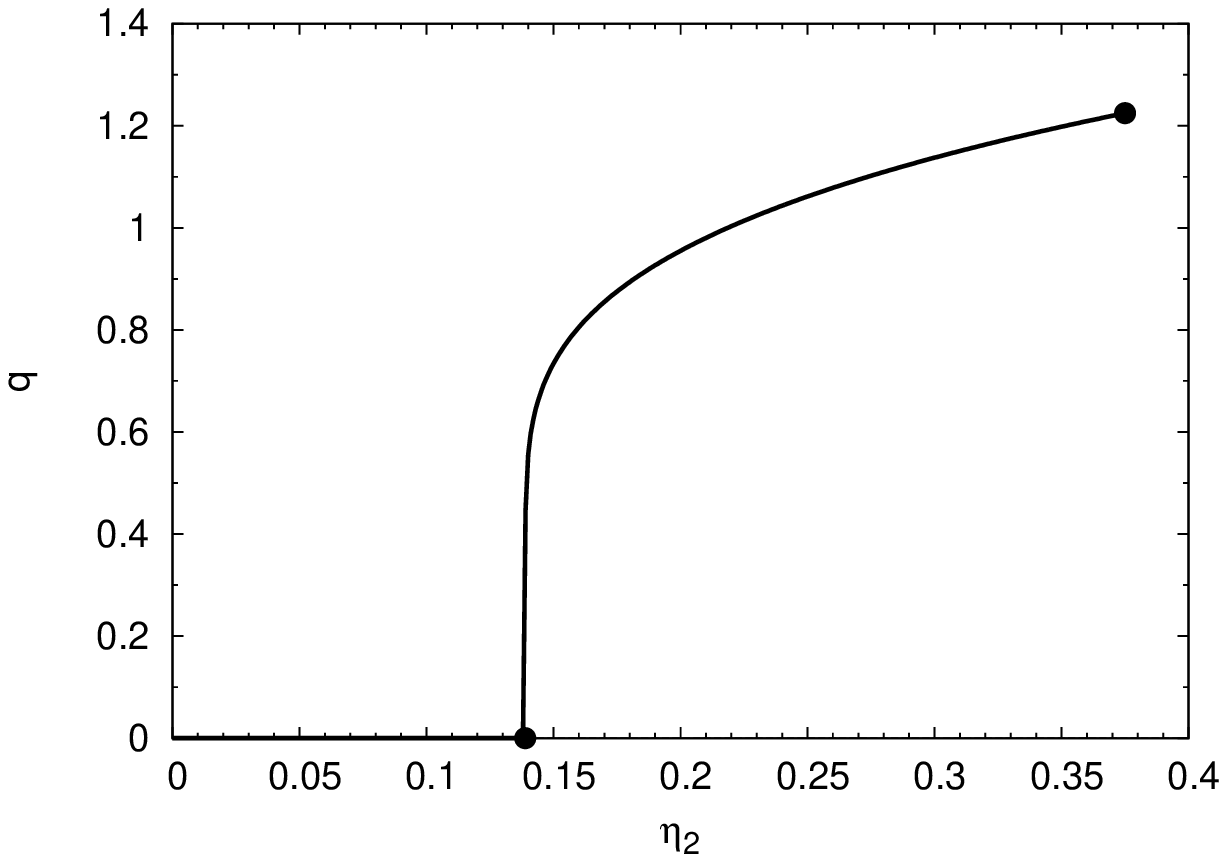}}
\caption{%
 (color online). {\bf (a)}~The amplitude of the dynamical mass and an
 effective order parameter $m_{\mathrm{ave}}$ in the solitonic phase as
 a function of $\eta_2$. 
 {\bf (b)}~The magnitude of wave vector
 $q={2\pi}/{\ell_{\mathrm{P}}}$ as a function of $\eta_2$ in the
 solitonic phase.}
\label{fig:snampperiod}
\end{figure}

We have seen that, for $\alpha_4<0$, the first-order chiral restoration at
$\alpha_2=\eta_2^{\mathrm{c}}\alpha_4^2/\alpha_6$ is replaced by two 
second-order critical lines, one at
$\alpha_2=\eta_2^{\mathrm{I}}\alpha_4^2/\alpha_6$
representing the onset of the formation of a single soliton in a
homogeneous sea of chiral condensate, and the other
at $\alpha_2=\eta_2^\mathrm{II}\alpha_4^2/\alpha_6$, corresponding to the 
chiral restoration from an inhomogeneous chiral phase.
In short, the chiral restoration becomes a smooth, hierarchical 
two-step transition.
So now the question is how does the system develop from a single soliton
state to a sinusoidal LO-type state as $\eta_2$ is
increased from $\eta_2^{\mathrm{I}}$ to $\eta_2^{\mathrm{II}}$?
This was already addressed in the paper
\cite{Buzdin:1997gg,Houzet:1999st,Basar:2008im,*Basar:2008ki}
where it is shown that the solution which covers these two situations
is the solitonic condensate characterized by Jacobi's elliptic function. 
Here we closely follow the discussion in \cite{Buzdin:1997gg} and
demonstrate that, if the modulation is restricted to 1D, Jacobi's
elliptic function gives in fact an adequate solution to the problem. 
Let us start with the Euler-Lagrange (EL) equation for the inhomogeneous
chiral condensate, 
$$
\frac{\delta}{\delta
M(\textbf{x})}\int_{\mathrm{WS}}d\textbf{y}\,\omega(M(\textbf{y}))=0.
$$ 
The condition results in the following fourth-order ordinary, but
nonlinear differential equation (when $\alpha_4<0$):
\beq
\ba{rcl}
 0&=&M^{(4)}(z)+3M''-10\left[M(M')^2+M^2M''\right]\\[1ex]
  & &+6\eta_2M-6M^3+6M^5.
\ea
\label{eq:E-L1}
\eeq
We need to solve this equation to find a suitable 1D modulation.
There is no systematic way to find out the general solution to a
fourth-order nonlinear differential equation. 
However it is not so difficult to see that Jacobi's elliptic
function is a particular solution to Eq.~\eqref{eq:E-L1}
\cite{Buzdin:1997gg}. 
Let $\mathrm{sn}(z,\nu)$ be Jacobi's elliptic function with $\nu$
being the elliptic modulus. We then set 
\beq
  M_{\mathrm{sn}}(z)=M_0\nu\,\mathrm{sn}(kz,\nu),
\label{eq:euler}
\eeq
where $k$ and $M_0$ are constants \footnote{%
In Ref.~\cite{Nickel:2009ke}, rather $\nu_2=\nu^2$ is treated as an argument
of Jacobi's elliptic function. In this notation Eq.~\eqref{eq:euler}
becomes $M_0\sqrt{\nu_2}\,\mathrm{sn}(kz,\nu_2)$.
}.
Following the discussion in \cite{Buzdin:1997gg}, we can show that
 this function obeys a fourth-order differential equation
\beq
\ba{rcl}
  0&=&M_{\mathrm{sn}}^{(4)}+(A+1)k^2(\nu^2+1)M_{\mathrm{sn}}''\\[2ex]
   & &-\frac{k^2}{M_0^2}(12-B)%
    \left[M_{\mathrm{sn}}(M_{\mathrm{sn}}')^2%
     +M_{\mathrm{sn}}^2M_{\mathrm{sn}}''\right]\\[1ex]
   & &+\left[Ak^4(\nu^2+1)^2-B\nu^2\right]%
      M_{\mathrm{sn}}\\[1ex]
   & &-\frac{2k^4}{M_0^2}(1+\nu^2)%
     \left(3-B+A\right)M_{\mathrm{sn}}^3%
     +\frac{3k^4}{M_0^4}\left(4-B\right)M_{\mathrm{sn}}^5,\\[2ex]
\ea
\label{eq:sndiff}
\eeq
with $A$ and $B$ being arbitrary numbers. Comparing Eq.~\eqref{eq:euler}
and Eq.~\eqref{eq:sndiff}, we see that $M_{\mathrm{sn}}$ constitutes a
sufficient solution to the EL equation \eqref{eq:euler} when the
following five algebraic equations are all satisfied:
\begin{subequations}
\begin{eqnarray}
  6\eta_2&=&\textstyle k^4\left[A(\nu^2+1)^2-B\nu^2\right],\label{eq:11a}\\[1ex]
  -3&=&\textstyle -(1+A)k^2(\nu^2+1),\label{eq:11b}\\[1ex]
  10&=&\textstyle \frac{k^2}{M_0^2}(12-B),\label{eq:11c}\\[1ex]
  -6&=&\textstyle -\frac{2k^4}{M_0^2}(1+\nu^2)(3-B+A),\label{eq:11d}\\[1ex]
  6&=&\textstyle \frac{3k^4}{M_0^4}(4-B).\label{eq:11e}
\end{eqnarray}
\end{subequations}
From Eqs.~\eqref{eq:11c} and \eqref{eq:11e}, we see $(M_0/k,B)=(1,2)$
or $(M_0/k,B)=(2,-28)$, 
but the latter results in complex $k^2$ and $\nu$. 
Then we take the choice $(M_0/k,B)=(1,2)$,
and in this case we see Eqs.~\eqref{eq:11b} and \eqref{eq:11d} are
degenerate so that we are left with only two algebraic equations
(\ref{eq:11a}) and (\ref{eq:11b}), whereas we still have
three unknown parameters, $\nu$, $k$, and $A$. 
From these two equations we can solve $\nu$ and $k$ as a function of
the continuous parameter $A$ (and $\eta_2$). 
Denoting these functions as $k_A$ and $\nu_A$, 
we arrive at a one-parameter solution group to the EL equation: 
\beq
  M_{\mathrm{sn}}(z;A)=k_A\nu_A\,\mathrm{sn}(k_Az,\nu_A).
\eeq
We stress that for any value of $A$, as long as $0\le\nu_A\le1$ and
$k_A^2>0$ are both satisfied, $M_{\mathrm{sn}}(z;A)$ gives a solution to
the EL equation.
The potential $\Omega(M_{\mathrm{sn}}(z;A))$ depends on $A$ so that we
need to look for an optimized value of $A$ demanding that it takes
a minimum.
This is easy to work out numerically 
for $\eta_2^{\mathrm{I}}\le\eta_2\le\eta_2^{\mathrm{II}}$.
We see $A\to 4/5$ as $\eta_2\to\eta_2^{\mathrm{I}}+0$, 
where $\nu_A$ approaches unity ($\nu_A\to 1$) and $k_A\to
k_{\mathrm{I}}=\sqrt{5/6}$,
\beq
\lim_{\eta_2\to\eta_2^{\mathrm{I}}+0}M_{\mathrm{sn}}(z;A)%
=k_{\mathrm{I}}\tanh(k_{\mathrm{I}}x),
\eeq 
while $A\to 1$ as the other end
$\eta_2=\eta_2^{\mathrm{II}}-0$ is approached, where $\nu_A$ goes
vanishing and $k_A\to k_{\mathrm{II}}\equiv\sqrt{3/2}$,
\beq
\lim_{\eta_2\to\eta_2^{\mathrm{II}}-0}M_{\mathrm{sn}}(z;A)/\nu_A%
=k_{\mathrm{II}}\sin(k_{\mathrm{II}}x).
\eeq
In the latter limit, $\nu_A$ goes zero so that the magnitude of
condensate $|M(z)|$ vanishes while its form becomes sinusoidal.
These two limits are given by exactly the same forms as we assumed in
the previous section. 
We show in Fig.~\ref{fig:snampperiod}(a) the amplitude of mass
function $\nu_A k_A$ as a function of $\eta_2$.
Also depicted together is an effective order parameter $m_{\mathrm{ave}}$ 
defined by the root-mean-square mass averaged over the elliptic
modulation period $\ell_{\mathrm{P}}=4K(\nu_A)/k_A$ with $K$ being the
complete elliptic integral of the first kind:
\beq
m_{\mathrm{ave}}=\sqrt{\langle M(z)^2\rangle}%
=k_A\sqrt{1-E(\nu_A)/K(\nu_A)}.
\eeq
$E$ is the complete elliptic integral of the second kind. 
We also show in Fig.~\ref{fig:snampperiod}(b) the magnitude of
wave vector $q=2\pi/\ell_{\mathrm{P}}$.
From these figures, we clearly see
that both phase transitions at
$\eta_2^{\mathrm{I}}$ and $\eta_2^{\mathrm{II}}$ are continuous ones.

\begin{figure}[tp]
\begin{center}
\includegraphics[scale=0.65]{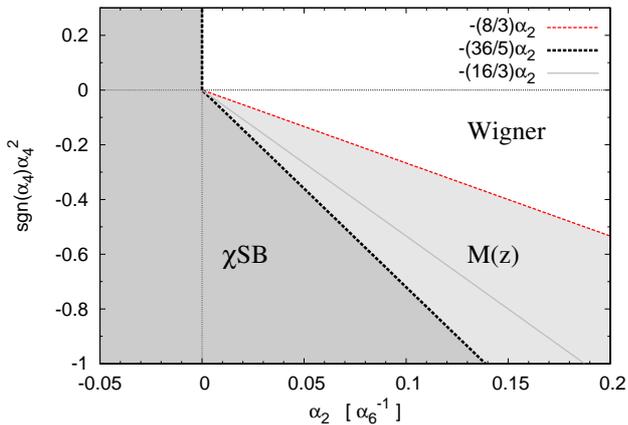}
\end{center}
\caption{(color online). The phase diagram in the
 $[\alpha_2,\sgn(\alpha_4)\alpha_4^2]$ plane. 
The horizontal axis represents $\alpha_2$ measured in the unit
 $[\alpha_6^{-1}]$, namely $\alpha_2\alpha_6$.
 For $\alpha_4\ge0$, the $\chi$SB phase ($\alpha_2<0$) and the 
 Wigner phase ($\alpha_2>0$) are separated at $\alpha_2=0$ by a 
 second-order phase transition. For $\alpha_4<0$, there are three phases:
 the homogeneous chiral condensate indicated by $\chi$SB, the solitonic
 chiral phase marked by $M(z)$, and the Wigner phase.
 Phase transitions separating these three phases, the dashed lines
 (black and red), are second order.
 For comparison, would-be first-order phase transition from the
 homogeneous $\chi$SB phase to the Wigner phase is also shown by the
 thin solid (gray) line inside the inhomogeneous phase.
}
\label{fig:phasediagram}
\end{figure}

Finally we show in Fig.~\ref{fig:phasediagram} the phase diagram in the
$[\alpha_2,\sgn(\alpha_4)\alpha_4^2]$ plane 
which is essentially the same as the one displayed in figure 1
of Ref.~\cite{Nickel:2009ke}.

\subsection{Singular behavior of thermodynamic quantities at the onset
  of soliton formation}
The phase transition at $\eta_2=\eta_2^{\mathrm{II}}$ from the Wigner
phase to the solitonic phase is second order, as clearly seen from the
GL potential Eq.~\eqref{eq:GLexforLO}. 
What about the order of transition for a soliton formation at
$\eta_2=\eta_2^{\mathrm{I}}$?
Near the critical point $\nu=1$, the modulation period diverges
logarithmically as $\ell_{\mathrm{P}}\sim 4K(\nu)\sim-2\log((1-\nu)/8)$
so that the spatial separation between domain walls,
a half period $\ell_{\mathrm{P}}/2$, also becomes large.
The corresponding wave vector $q={2\pi}/{\ell_{\mathrm{P}}}%
=\frac{\pi k}{2K(\nu)}$ behaves as
\beq
\textstyle
  q(\nu\to 1^{-})=-\frac{\pi\sqrt{5/6}}{\log((1-\nu^2)/16)}.
\eeq
We see that the wave vector drops to zero, inversely proportional to the
logarithm of $(1-\nu)$, just as quark number density
in the GN model \cite{Schnetz:2005ih,Schnetz:2004vr}.
The problem is how the elliptic modulus behaves as a function of
$\eta_2$ near the critical point. As $\eta_2^{\mathrm{I}}$ is
approached, we see $\nu\to 1$, $k^2\to 5/6$ with $A\to 4/5$. 
From Eqs.~(\ref{eq:11a}) and (\ref{eq:11b}),
we can extract the expansion of $\nu$ and $k^2$ 
in terms of $(4/5-A)$ and $\eta_2-\eta_2^{\mathrm{I}}$. 
To first nontrivial order, we find
\beq
\ba{rcl}
 1-\nu^2&=&\frac{4}{\sqrt{3}}\sqrt{(\frac{4}{5}-A)+\frac{81}{25}(\eta_2-\eta_2^\mathrm{I})},\\[2ex]
 k^2-\frac{5}{6}&=&\frac{5}{3\sqrt{3}}\sqrt{(\frac{4}{5}-A)+\frac{81}{25}(\eta_2-\eta_2^\mathrm{I})}.
\ea
\eeq
Also the relation between $A$ and $\eta_2-\eta_2^{\mathrm{I}}$ was
numerically found to be linear in the vicinity of critical point
($\eta_2\agt\eta_2^{\mathrm{I}}$), as
$A\cong\frac{4}{5}-2.406(\eta_2-\eta_2^{\mathrm{I}})$. 
We obtain
\beq
\ba{rcl}
 1-\nu^2&=&\frac{2}{\sqrt{3}}\kappa_{\mathrm{I}}\sqrt{\eta_2-\eta_2^\mathrm{I}},\\[2ex]
 k^2-\frac{5}{6}&=&\frac{5}{3\sqrt{3}}\kappa_{\mathrm{I}}\sqrt{\eta_2-\eta_2^\mathrm{I}},
\ea
\eeq
with a numerical value of $\kappa_{\mathrm{I}}\cong 2.38$.
Therefore, when the critical
point $\eta_2^{\mathrm{I}}$ is approached from above, the inverse of
the modulation length behaves as
\beq
 q(\eta_2\to\eta_2^{\mathrm{I}}+0)%
  \sim-\frac{\pi\sqrt{{10}/{3}}}{\log\left(\frac{\kappa_{\mathrm{I}}^2}{48}%
  \left(\eta_2-\eta_2^{\mathrm{I}}\right)\right)}.
\eeq
We have checked to see that this analytical formula excellently
reproduces the numerical result displayed in Fig.~\ref{fig:snampperiod}(b).
This means that the separation of domain walls becomes logarithmically
large while the wall thickness (the size of defect $\sim1/k$) does not
change drastically near $\eta_2^{\mathrm{I}}$.

We now look at the singular behaviors in the thermodynamic quantities.
These are determined by the behavior of derivatives of the GL free
energy with respect to the parameter $\eta_2$. 
The first derivative in the vicinity of the critical point can be computed up
to the first few orders in the expansion in $\eta_2$ as
\beq
\ba{rcl}
 \frac{\partial\Omega}{\partial\eta_2}&=&\frac{1}{2}\langle
 M(z)^2\rangle_{\mathrm{WS}}\\[2ex]
 &\cong&\left\{\begin{array}{lcc}
  \frac{5}{12}-\frac{3}{4}(\eta_2-\eta_2^{\mathrm{I}})+\cdots,&
   &\mbox{$(\eta_2<\eta_2^{\mathrm{I}})$}\\[2ex]
 \frac{5}{12}+\frac{5}{3\log\big(\frac{\kappa_{\mathrm{I}}^2}{48}%
  (\eta_2-\eta_2^{\mathrm{I}})\big)}+\cdots.&&\mbox{$(\eta_2>\eta_2^{\mathrm{I}})$}
  \end{array}
 \right.
\ea
\eeq
We see that the first derivative is continuous. 
The second derivative is also worked out, and the first two dominant parts
(either regular or singular) can be extracted as
\beq
 \textstyle\frac{\partial^2\Omega}{\partial\eta_2^2}%
  \sim\left\{\begin{array}{lcc}
  -\frac{3}{4}-\frac{27}{8}(\eta_2-\eta_2^{\mathrm{I}})+\cdots,&
   &\mbox{$(\eta_2<\eta_2^{\mathrm{I}})$}\\[2ex]
 -\frac{5+\frac{5}{3}\kappa_{\mathrm{I}}%
  \sqrt{\eta_2-\eta_2^{\mathrm{I}}}}%
  {3(\eta_2-\eta_2^{\mathrm{I}})\log^2\big(\frac{\kappa_\mathrm{I}^2}{48}%
  (\eta_2-\eta_2^\mathrm{I})\big)}+\cdots.
  &&\mbox{$(\eta_2>\eta_2^{\mathrm{I}})$}
  \end{array}
 \right.
\label{eq:suscep}
\eeq
It is clear that the second derivative is not only discontinuous but
diverges as the critical point is approached from the side of
the inhomogeneous phase. The singular behavior is characterized mainly 
by the power law but with a logarithmic correction just as with the
fermion number susceptibility in the GN model
\cite{Schnetz:2005ih,Schnetz:2004vr}. 
We see that the transition is second order. 
We remark that the same singularity was obtained also in the NJL
model \cite{Carignano:2010ac} where it was shown further
that the vector-type interaction between quarks washes out the singular
behavior.

Before closing this section, we briefly discuss the origin of the
singular behavior of the second derivative of $\Omega$. 
At $T=0$ the singularity is related to that of quark number
susceptibility \cite{Schnetz:2005ih,Schnetz:2004vr,Carignano:2010ac}.
To see this, we first note that at $T=0$ a net quark number
(imbalance to $\bar{q}q$ condensate) tends to accumulate in the wall
surfaces where the magnitude of the chiral condensate vanishes. 
Then considering the fact that the separation length between domain walls
diverges as $\ell_{\mathrm{P}}/2\sim 1/\log(\eta_2-\eta_2^{\mathrm{I}})$ while the
size of the wall where quarks are present, $1/k$, stays almost constant, 
the averaged quark number density may be approximately computed as
\beq
 {\langle q^\dagger q\rangle}%
 \sim n_0\frac{1/k}{\ell_{\mathrm{P}}+1/k}%
 \sim-\frac{n_0}{\log\left(\frac{\kappa_{\mathrm{I}}}{48}%
  (\eta_2-\eta_2^{\mathrm{I}})\right)},
\eeq
with $n_0$ defined as the quark density in the absence of
a condensate.
Assuming $n_0$ does not change significantly in the vicinity of
the critical point, the quark number susceptibility on the side
of the inhomogeneous phase becomes
\beq
  \frac{d\langle q^\dagger q\rangle}{d\mu}%
  \sim\frac{d\eta_2}{d\mu}\frac{n_0}{(\eta_2-\eta_2^{\mathrm{I}})%
  \left[\log\left(%
  \frac{\kappa_{\mathrm{I}}}{48}(\eta_2-\eta_2^{\mathrm{I}})\right)\right]^2}.
\eeq
This has exactly the same parametric dependence on $\eta_2$ 
as Eq.~\eqref{eq:suscep}, suggesting that at $T=0$ the divergence of the
second derivative of $\Omega$ has originated in the divergent quark
number susceptibility \cite{Schnetz:2005ih,Schnetz:2004vr,Carignano:2010ac}.

\section{Other crystallization patterns in 1D and higher 
dimensional  modulations}\label{sec:multi}
In this section we address the question of whether or not states other
than the solitonic condensate, with either 1D or higher dimensional
modulations, are possible. 
We work near the Lifshitz point, so we retain up to sixth order in
GL expansion as in the previous section. 
In Sec.~\ref{sec21}, we discuss the thermodynamics of the LO-like chiral
condensate, Eq.~\eqref{eq:sin}, and the FF-like
chiral spiral \cite{Nakano:2004cd} in some detail. 
Even though these two states are less favorable than the solitonic
solution, the analysis still serves as an illustrative
benchmark when we explore higher dimensional modulations. 
In Sec.~\ref{sec21}, we try a general {\t Ansatz} for the 1D
modulation, \ie, the condensate expanded in higher harmonics. 
For comparison, we also make a harmonic analysis on the solitonic
state.
In Sec.~\ref{sec22} we introduce higher dimensional analogs of the
LO-like chiral density wave, and see if such higher
dimensional chiral lattices can be realized near the Lifshitz
point.

\subsection{Real chiral density wave and chiral spiral}%
\begin{figure}[t]
\centering
\subfigure[]{\includegraphics[scale=0.6]{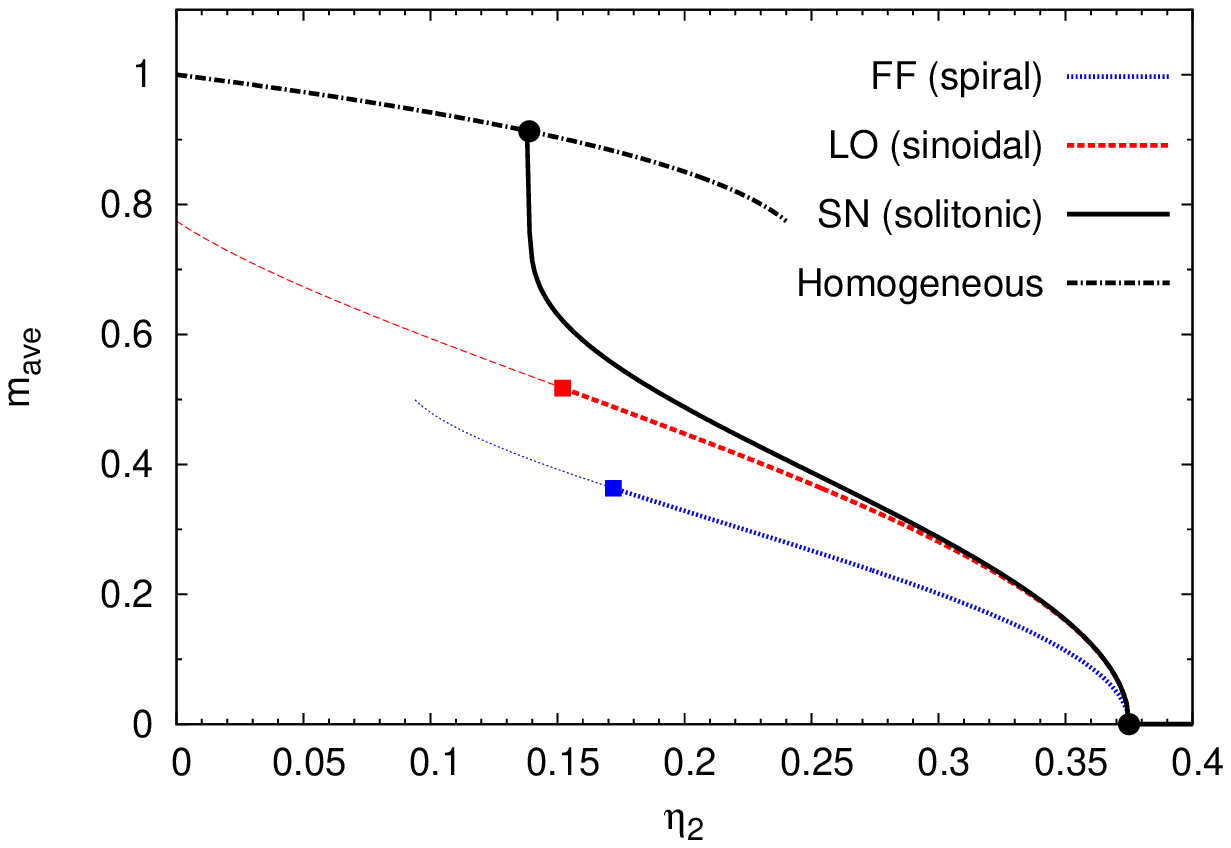}}
\vspace*{-1ex}
\subfigure[]{\includegraphics[scale=0.6]{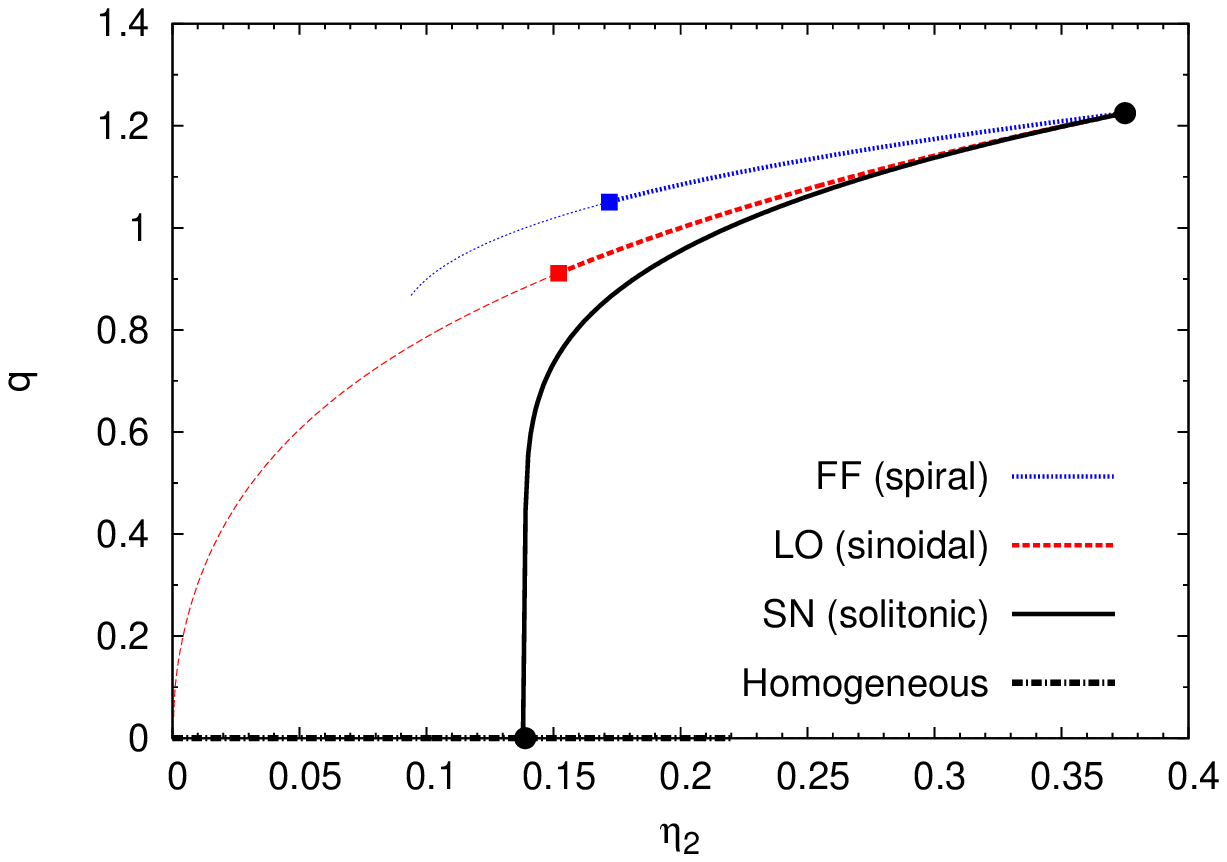}}
\hspace*{-3.1ex}
\subfigure[]{\includegraphics[scale=0.63]{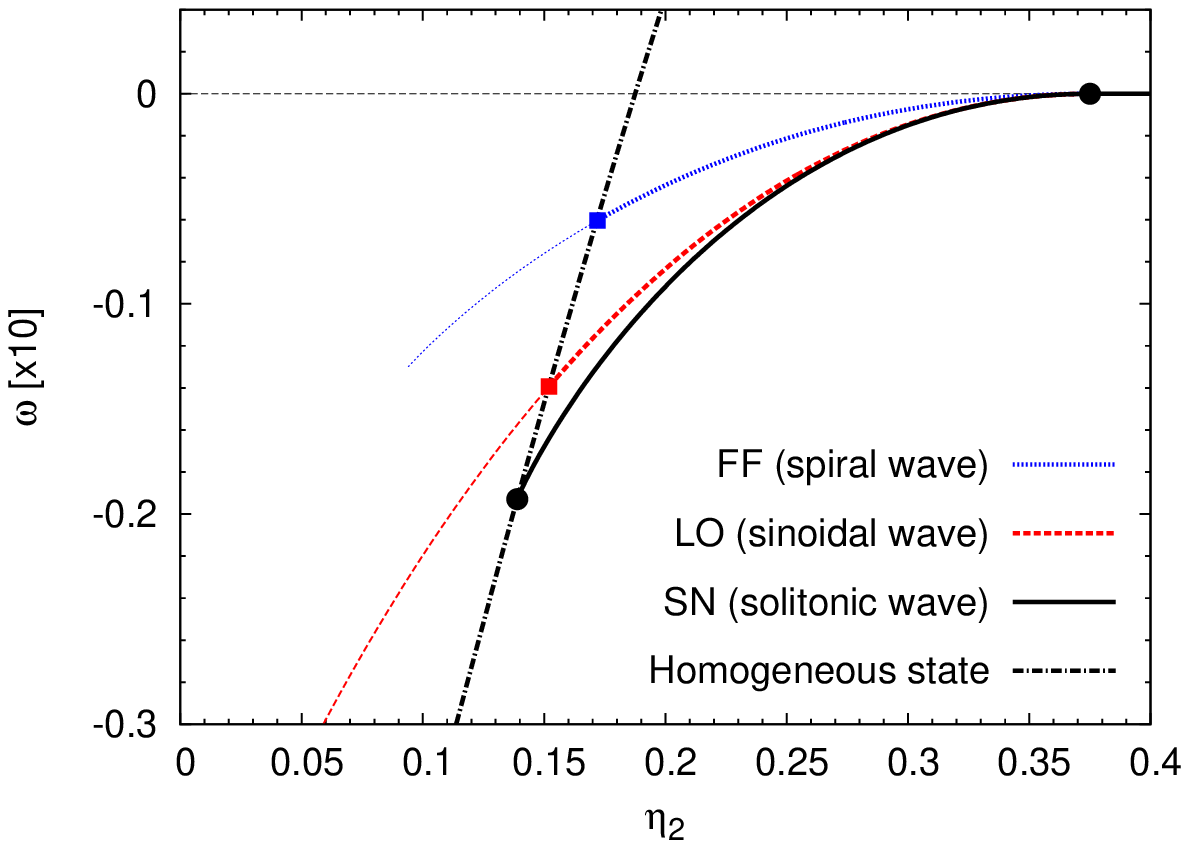}}
\caption{(color online). {\bf (a)}~Effective order parameters, 
 {\bf (b)}~wave vectors, and {\bf (c)}~potentials as a
 function of $\eta_2$ for the LO (dashed, red), FF
 (dotted, blue) and SN (solid) states.
 The quantities corresponding to the homogeneous condensate are shown by
 dot-dashed lines.
 The solid lines are the same as those depicted in Fig.~\ref{fig:snampperiod}.
}
\label{fig:FFLOcomp}
\end{figure}

\label{sec21}
Here we discuss the thermodynamics of two typical chiral density
waves: a real, LO-like sinusoidal chiral density wave of the form
Eq.~\eqref{eq:sin} and a FF-like chiral spiral characterized by a single
plane wave  
\beq
  M_{\mathrm{FF}}(z)=M_0e^{ikz}.
\eeq
The imaginary part should be understood as the pseudoscalar condensate,
for one choice in the charge neutral channel:
$M_{\mathrm{FF}}(\textbf{x})\sim-2G(\langle\bar{q}q\rangle +
i\langle\bar{q}\gamma_5i\tau_3q\rangle)$.
This is also called the ``dual chiral density wave'' abbreviated to the
DCDW in the original paper \cite{Nakano:2004cd}.
In the following we refer to this state mainly as the FF state or chiral
spiral.
When the chiral condensate involves a finite imaginary part, we need to
generalize the GL functional so as to allow it. 
The generalized functional for complex chiral condensates but with
restriction to 1D modulations was worked out first in the chiral GN
model
\cite{Boehmer:2007ea}; the resulting functional has terms asymmetric
with respect to an interchange between the real and imaginary parts of
the condensate, $\re(m)\leftrightarrow\im(m)$. 
These terms are responsible for the realization of the FF-type complex
chiral condensate in the GN model \cite{Schon:2000qy} and also in the
$(1+1)$-dimensional NJL model in the large $N$ limit
\cite{Basar:2009fg,Ebert:2011rg}.
In the NJL model in the 1+3 dimension, however, these terms are shown to
vanish \cite{Nickel:2009wj}.
Then the functional is cast in the following form in the chiral limit,
using the same convention as Eq.~\eqref{eq:omega}:
\beq
\ba{rcl}
 \tilde{\omega}&=&\frac{\eta_2}{2}|m|^2+\frac{1}{4}{\sgn(\alpha_4)}%
  (|m|^4+|m'|^2)\\[1ex]
  &&\!\!\!\!+\frac{1}{6}\big(|m|^6+4|m|^2|m'|^2%
+\re(m')^2(m^*)^2+\frac{1}{2}|m''|^2\big).\\[1ex]
\ea
\eeq
When the condensate is real [$m(z)^*=m(z)$], this functional reduces to
Eq.~\eqref{eq:omega} with the restriction to the 1D modulation,
$M(\textbf{x})\to M(z)$, substituted. 
As before, we concentrate on the case $\alpha_4<0$. 
In Fig.~\ref{fig:FFLOcomp} we show the amplitude of masses~(a),
the magnitude of wave vectors~(b),
and corresponding thermodynamic potentials~(c) for the 
FF and LO condensates, as a function of $\eta_2$.
The wave vector $q$ is just $k$ ($k_{\mathrm{II}}$) for the FF (LO)
state, while for the solitonic state it is given by
$2\pi/\ell_{\mathrm{P}}$.
From the figures, we see that the real LO phase is more favorable than
the FF-like chiral spiral over the whole range of $\eta_2$, 
while they are less favorable than the solitonic phase, which will
often be denoted by the solitonic (SN) state hereafter.
The FF and LO (and SN) states become degenerate at the onset of
condensate at $\eta_2^{\mathrm{II}}=3/8$.
Near the point it is easy to perform the expansion in $M_0$. 
The result for the LO phase is already given by
Eq.~\eqref{eq:GLexforLO}, and that for the FF phase is
obtained at the same order in $M_0$ as \cite{Nickel:2009wj}
\beq
\ba{rcl}
  \Omega&=&\langle\omega(M_{\mathrm{FF}}(\textbf{x}))%
  \rangle_{\mathrm{WS}}\\[1ex]
  &=&\left(\frac{\eta_2}{2}-\frac{3}{16}\right)M_0^2%
  +\frac{1}{2}M_0^4+{\mathcal O}(M_0^6).
\ea
\eeq
The coefficient of $M_0^4$ for the FF phase is twice as large as
that of the LO phase, from which we see that the energy for the FF phase
is higher than that for the LO phase at least near the onset.
This may be reasonably understood along the same line as 
\cite{YoshidaYip,Bulgac:2008tm} where 
it is stressed that additional terms are required in the effective
theory in order to recover the broken time reversal symmetry in the FF
state, which results in an extra energy cost.

Since the LO phase is more favorable than the FF phase, we concentrate
on the LO phase in the following. 
The transition between the homogeneous $\chi$SB and LO phases is of 
first order. 
We can obtain the critical $\eta_2$ by solving
$\min_{M}\omega(M_{\mathrm{const.}})%
=\min_{M,k}\langle\omega(M_0\sin
kz)\rangle_{\mathrm{WS}}$. 
This was worked out numerically and the location of the critical point
was found as
\beq
\eta_2({\mathrm{LO}\leftrightarrow\chi\mathrm{SB}})\cong0.1519,
\eeq
which is larger than $\eta_2^{\mathrm{I}}\sim 0.1389$ for a soliton
formation, reflecting the fact that the LO phase is metastable
against the SN phase. 
Crossing the critical point from the $\chi$SB phase to the LO phase, the
magnitude of the condensate drops by about 20\%. 
Also the ratio of the amplitude of mass to the magnitude of the
wave vector in the LO phase just at the critical point has been found
to be about 0.8. 
These can be summarized as
\beq
 \frac{M_0(\mathrm{LO})}{M_{\mathrm{const.}}(\chi\mathrm{SB})}%
 \cong0.81,\quad\frac{M_0(\mathrm{LO})}{k(\mathrm{LO})}\cong0.80.
\eeq
We note that all these ratios are the universal constants associated with
the first-order phase transition between the $\chi$SB and LO phases, 
at the sixth order we are working.

We also remark that the first {\it Ansatz} of counting $M_0$ and $k$ as the
same order in magnitude is consistent at this transition. 
At the onset of the condensate, however, the amplitude $M_0$ vanishes while
$k$ remains finite so that the derivative terms are more
important than the bulk homogeneous terms. 
In contrast, at the onset of the domain wall formation, the derivative
terms play a minor role.

\subsection{Most general condensate with higher harmonics}\label{sec22}
We here try to see if Jacobi's elliptic function is the most
favorable solution among 1D modulation patterns.
We first note that Jacobi's elliptic function just gives a one
parameter subgroup of solutions to the EL equation. 
It should be stressed that the original EL equation \eqref{eq:E-L1} 
is a nonlinear fourth-order differential equation
while Eq.~\eqref{eq:sndiff} is basically the sum of three differential
equations each of which can be obtained from the second order
differential equation which the elliptic function obeys.
It is thus not obvious that it really covers all the solutions to
the original EL equation. 
Keeping this in mind, we try the most general assumption for the
spatially modulated chiral condensate. 
We set the condensate in the form of a harmonic expansion series as
\beq
  M_{\mathrm{HH}}(z;\{\ell_{\mathrm{H}},M_n^{\mathrm{H}}\})%
  =\sum_{n=1,3,5,\cdots}M_n^{\mathrm{H}}\sin((2\pi/\ell_{\mathrm{H}})nz),
\label{eq:HH}
\eeq
where $\{M_n\}$ and $\ell_{\mathrm{H}}$ are variational parameters.
We note that the even components are absent if we restrict
the condensate to a half antiperiodic period
$M(z+\ell_{\mathrm{H}}/2)=-M(z)$. 
This is quite reasonable for the case of the chiral limit. In fact we
have checked numerically that the even components are vanishing even if
included.
Hereafter, we abbreviate the state characterized by $M_{\mathrm{HH}}$ as
the higher harmonic (HH) state.
We took into account up to the fifth harmonics in the expansion above
and minimized the thermodynamic potential with respect to
$M_1^{\mathrm{H}},M_3^{\mathrm{H}},M_5^{\mathrm{H}}$ and the modulation
period $\ell_{\mathrm{H}}$. 

\begin{figure}[tbp]
\centering
\subfigure[]{\includegraphics[scale=0.55]{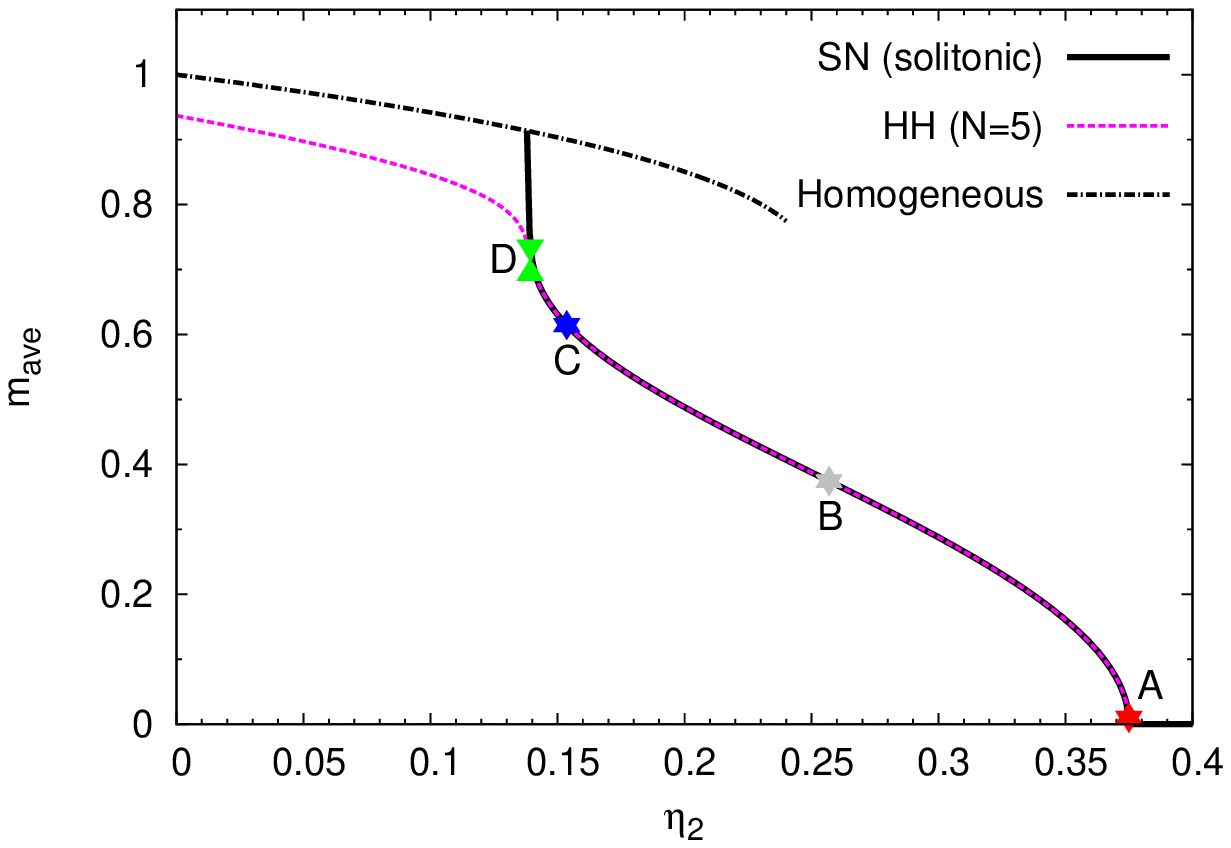}}
\subfigure[]{\includegraphics[scale=0.57]{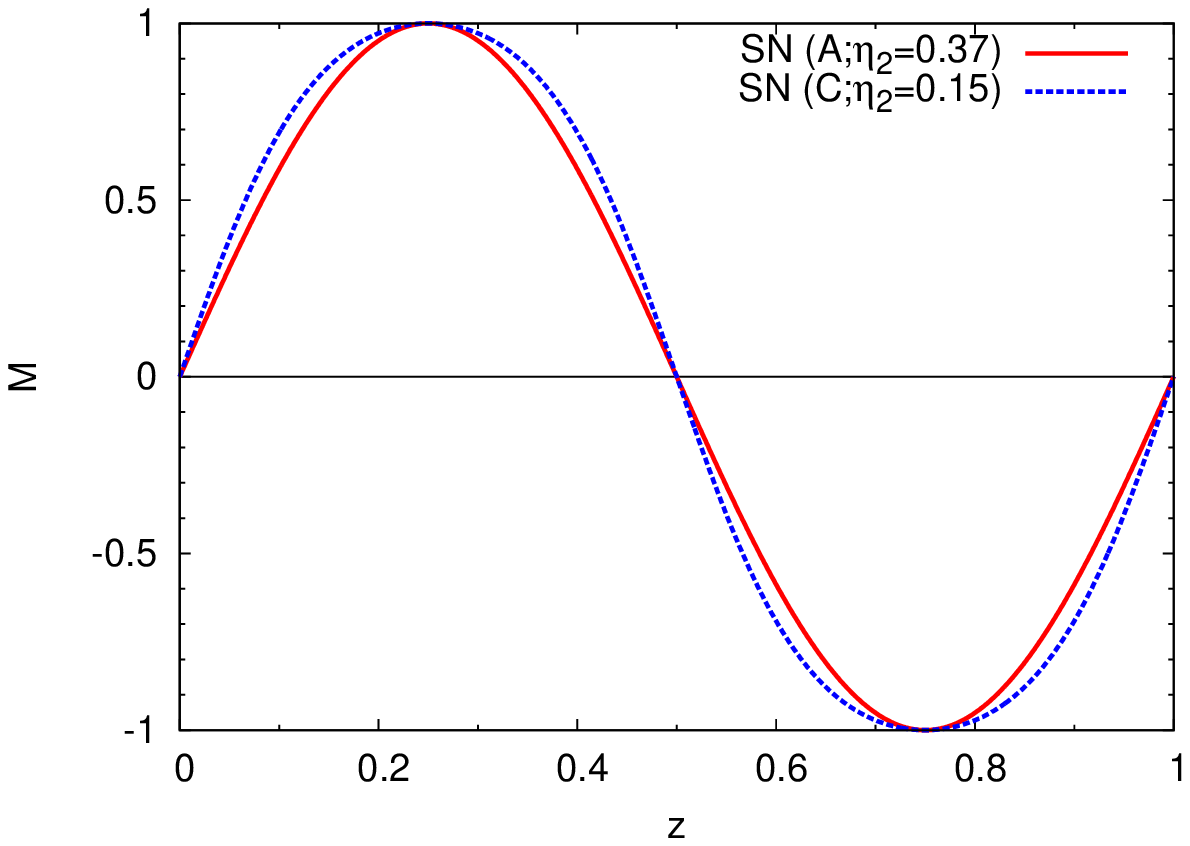}}
\subfigure[]{\includegraphics[scale=0.57]{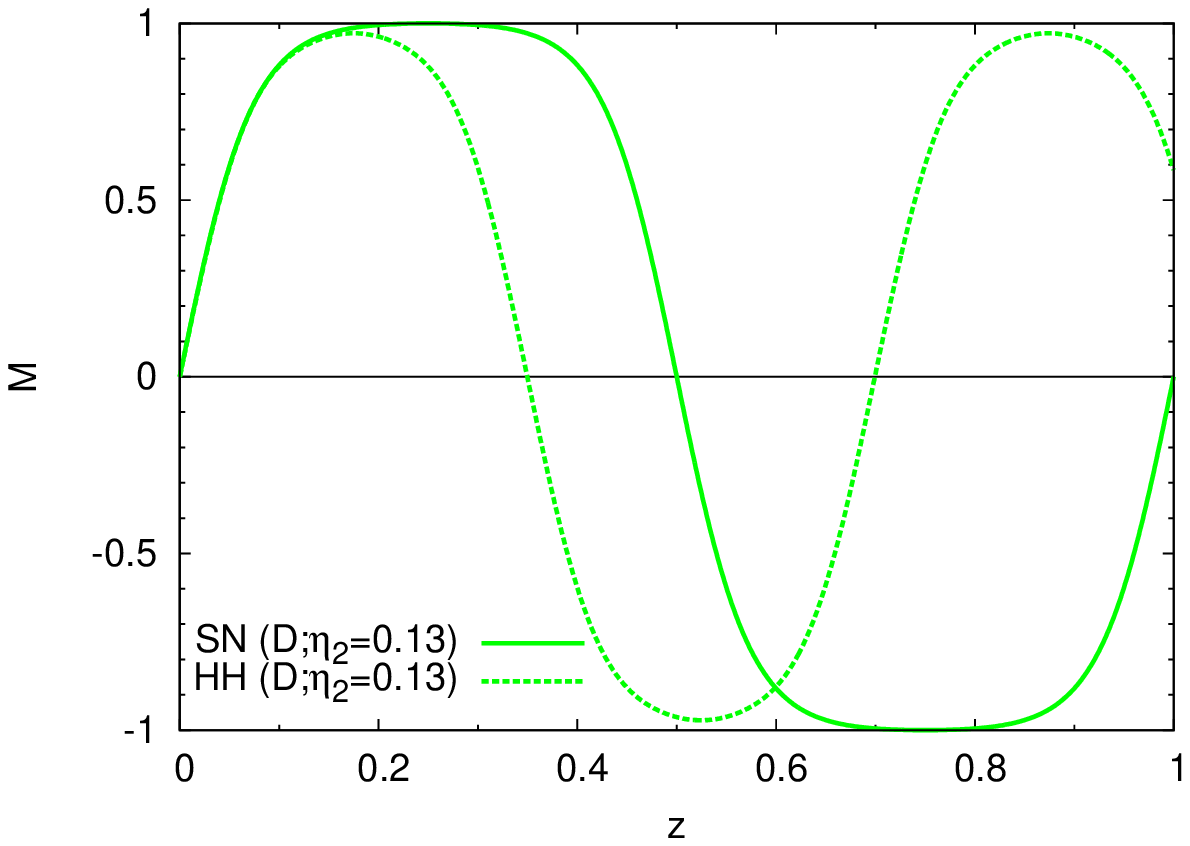}}
\caption{%
(color online). {\bf (a)}~Spatial profile of the condensate
 $M_{\mathrm{sn}}(z)$ at
 points {\bf A} ($\eta_2=0.3759$; solid) and {\bf C} ($\eta_2=0.1536$;
 dashed). For each case, the difference from $M_{\mathrm{HH}}(z)$ is
 invisible so it is suppressed from being depicted.
 The unit of horizontal (vertical) axis $z$ ($M$) is set to the elliptic
 period (amplitude) of $M_{\mathrm{sn}}(z)$, \ie,
 $\ell_{\mathrm{P}}(\eta_2)$ ($k(\eta_2)\nu(\eta_2)$) at each point. 
{\bf (b)}~$M_{\mathrm{sn}}$ (solid) and $M_{\mathrm{HH}}$ (dashed) at point
 {\bf D} ($\eta_2=0.1389$). 
 The scales for $M$ and the spatial coordinate $z$ are normalized by the
 amplitude and the period of $M_{\mathrm{sn}}(z)$ as in (a).}
\label{fig:comp_profile}
\end{figure}

The resulting energy of the HH state appeared to be very close to that
of the SN state.
In order to illustrate the fact, we depict in
Fig.~\ref{fig:comp_profile}(a) the effective order parameter
$m_{\mathrm{ave}}=\sqrt{\langle M_{\mathrm{HH/SN}}(z)^2\rangle}$ as a
function of $\eta_2$, both for the SN and HH states: the solid line for the
SN state, and the dashed (magenta) line for the HH state.
We have labeled four representative points, $\eta_2=0.3749$, $0.2569$,
$0.1536$, and $0.1389$, by letters {\bf A}, {\bf B}, {\bf C}, and {\bf D}.
In the figure, the coordinate $(\eta_2,m_{\mathrm{ave}})$ at each point
for the SN (HH) state is marked by the lower (upper) triangle.
We see that the difference between the HH and SN states is visible
only at point {\bf D} which is very close to $\eta_{\mathrm{I}}$ for the
domain wall onset.

The spatial profile of the HH-condensate, $M_{\mathrm{HH}}(z)$, was also
observed to converge to that of the solitonic state, $M_{\mathrm{sn}}(z)$;
the difference is actually invisible already at this level of truncation
in the biggest range of $\eta_2$. 
It only becomes slightly manifest in the very vicinity of
$\eta_2=\eta_2^{\mathrm{I}}$ where the condensate becomes a
collection of domain wall solitons.
In Figs.~\ref{fig:comp_profile}(b) and (c), we show the spatial profiles
of the mass function at points {\bf A}, {\bf C} and {\bf D}.
Only at {\bf D} do we show both $M_{\mathrm{HH}}$ and $M_{\mathrm{SN}}$,
since at other points the differences are invisible.

\begingroup
\begin{table*}[t]
\caption{Comparison between $M_{\mathrm{sn}}(z)$ and
 $M_{\mathrm{HH}}(z)$.
The energy densities $\Omega_{\mathrm{sn}}$ and $\Omega_{\mathrm{HH}}$,
the values of Fourier strengths
$\{M_1^{\mathrm{H/sn}},M_3^{\mathrm{H/sn}},M_5^{\mathrm{H/sn}}\}$, and
 modulation periods $\ell_{\mathrm{H}}$ and $\ell_{\mathrm{P}}(\nu,k)$
 are tabulated at four representative points
$\eta_2=0.375,\,0.257,\,0.154$, and $0.1393$ which are assigned by the
 letters {\bf A}, {\bf B}, {\bf C}, and {\bf D}, respectively.}
\label{tab:sn1}
\centering
\begin{tabular}{|c||r||r|r||r|r||r|r||r|r||r|r||}
 \hline
\multicolumn{1}{|c||}{}&
\multicolumn{1}{c||}{$\nu$} & 
\multicolumn{1}{c|}{$\ell_{\mathrm{P}}$} &
\multicolumn{1}{c||}{$\ell_{\mathrm{H}}$} &
\multicolumn{1}{c|}{$\Omega_{\mathrm{sn}}$} & 
\multicolumn{1}{c||}{$\Omega_{\mathrm{HH}}$} &
\multicolumn{1}{c|}{$M_{1}^{\mathrm{sn}}$} &
\multicolumn{1}{c||}{$M_{1}^{\mathrm{H}}$} &
\multicolumn{1}{c|}{$M_{3}^{\mathrm{sn}}$} &
\multicolumn{1}{c||}{$M_{3}^{\mathrm{H}}$} &
\multicolumn{1}{c|}{$M_{5}^{\mathrm{sn}}$} &
\multicolumn{1}{c||}{$M_{5}^{\mathrm{H}}$}\\ \hline%
 {\bf A} $(\eta_2=0.3749)$ & $0.009$ & $5.13$ & $5.13$ & $-9\times
		     10^{-10}$ & 
            $-9\times10^{-10}$ & $0.0111$ & $0.0111$ & $6\times10^{-8}$
				 & $6\times10^{-8}$ &
	    ${\mathcal O}(10^{-13})$ & ${\mathcal O}(10^{-13})$
            \\ \hline
 {\bf B} $(\eta_2=0.2569)$ & $0.458$ & $5.85$ & $5.85$ & $-0.0038$ & 
            $-0.0038$ & $0.5291$ & $0.5291$ & $0.0077$ &
            $0.0077$ & $0.0001$ & $0.0001$\\ \hline
 {\bf C} $(\eta_2=0.1536)$ & $0.821$ & $8.22$ & $8.21$ &
		     $-0.0161$ & $-0.0161$ & $0.8673$ & $0.8665$ &
				 $0.0563$ & 
            $0.0559$ & $0.0039$ & $0.0038$
           \\ \hline
 {\bf D} $(\eta_2=0.1389)$ & $0.991$ & $14.98$ & $10.48$ &
		     $-0.0193$ & $-0.0192$ & $1.0672$ & $0.9781$ &
				 $0.1951$ & $0.1077$ & $0.0455$ &
					     $0.0132$ \\ \hline
\end{tabular}
\end{table*}
\endgroup

In order to see most efficiently how the HH and SN states diverge 
as the soliton onset is approached, we now make the harmonic analysis on
the solitonic condensate $M_{\mathrm{sn}}$. 
First we note that the decomposition of the elliptic function in the 
Fourier-Lambert series is given as
\beq
\ba{rcl}
 M_{\mathrm{sn}}(z;\{\nu,k\})&=&\disp\sum_{n=1,3,\cdots}^{\infty}%
 M_n^{\mathrm{sn}}(\nu,k)\sin\left(\frac{2\pi
 nz}{\ell_\mathrm{P}(\nu,k)}\right),\\[1ex]
 M_n^{\mathrm{sn}}(\nu,k)&=&\disp\frac{2}{\ell_{\mathrm{P}}(\nu,k)}\frac{4\pi
 Q(\nu)^{n/2}}%
{1-Q(\nu)^n},
\ea
\eeq
where $\ell_{\mathrm{P}}(\nu,k)$ is the elliptic (real) period given by
an explicit form $\ell_{\mathrm{P}}(\nu,k)={4K(\nu)}/{k}$, and
$Q(\nu)$ is the nome defined by
$Q(\nu)=\exp\left({-\pi{K'(\nu)}/{K(\nu)}}\right)$
with $K(\nu)$ and $iK'(\nu)\equiv
iK(\sqrt{1-\nu^2})$ being the quarter periods of Jacobian elliptic
functions.
We compare $\{M_n^{\mathrm{sn}}(\nu,k), \ell_{\mathrm{P}}(\nu,k)\}$
computed from $M_{\mathrm{sn}}(z)$ with
$\{M_n^{\mathrm{H}},\ell_\mathrm{H}\}$.
In Table.~\ref{tab:sn1}, we tabulate the values of potential, 
elliptic modulus, and harmonic coefficients computed from the
SN state $M_{\mathrm{sn}}$ and those for the HH state
at four points introduced above, {\bf A}, {\bf B}, {\bf C}, and 
{\bf D}.
From the table, we see that the energies of these two condensates are 
nearly degenerate, actually the difference is visible only at the point
{\bf D} located in the vicinity of the onset of soliton formation. 
Nevertheless it is true that the energy for the solitonic SN state
is always smaller than that for the HH state.
This means that even if we start from the general {\it Ansatz}
$M_{\mathrm{HH}}(z)$, the optimized spatial profile of
$M_{\mathrm{HH}}(z)$ approaches that of $M_{\mathrm{sn}}(z)$. 
This fact suggests that the elliptic function gives the absolute ground
state.
We recognize that the closer to the point $\eta_2^{\mathrm{I}}$ we get,
the more relevant the higher harmonic components become.
Near $\eta_2^{\mathrm{I}}$, $M_{\mathrm{HH}}(z)$ starts to fail to
approximate $M_{\mathrm{sn}}(z)$; the departure of $M_{\mathrm{HH}}(z)$
from $M_{\mathrm{sn}}(z)$ simply reflects the fact
that truncated higher harmonic components $(n>5)$ can no longer be
 neglected.
This is what actually is seen in Fig.~\ref{fig:comp_profile}(b) and (c);
$M_{\mathrm{HH}}(z)$ at points {\bf A} and {\bf C} well approximates
$M_{\mathrm{sn}}(z)$, while it underestimates both the amplitude and
the period at point {\bf D}.

\subsection{Multidimensional modulation}\label{sec23}
We now explore the possibility of realization of higher
dimensional crystals favored over the 1D solitonic state. 
We remark that the EL equation (\ref{eq:E-L1}) was obtained after the
restriction of condensate to one varying in 1D.
The original EL equation in 3D is
\beq
 \ba{rcl}
 0&=&\Delta^2 M({\textbf x})+3\Delta M({\textbf x})%
     -10\left[M(\nabla M)^2+M^2\Delta M\right]\\[1ex]
  & &+6\eta_2M-6M^3+6M^5,
\ea
\label{eq:E-L3D}
\eeq
where the operator $\Delta=\partial_x^2+\partial_y^2+\partial_z^2$ is
the three-dimensional Laplacian. 
This is a fourth-order nonlinear {\it partial} differential equation.
The solution space for the partial differential equation is much wider
than that for the ordinary differential equation. 
Therefore it is not trivial that Jacobi's elliptic function stays
as the most favorable structure when the restriction to 1D modulations is
taken away.
Here we do not search for the formal solution, but only try some
specific crystals having the simple square or cubic symmetry.
In order to demonstrate how the dimensionality of a crystal structure
affects the free energy, we concentrate on the LO-type phase which
has a simpler form than the elliptic function.
\begin{figure}[tbp]
\centering
\subfigure[]{\includegraphics[scale=0.6]{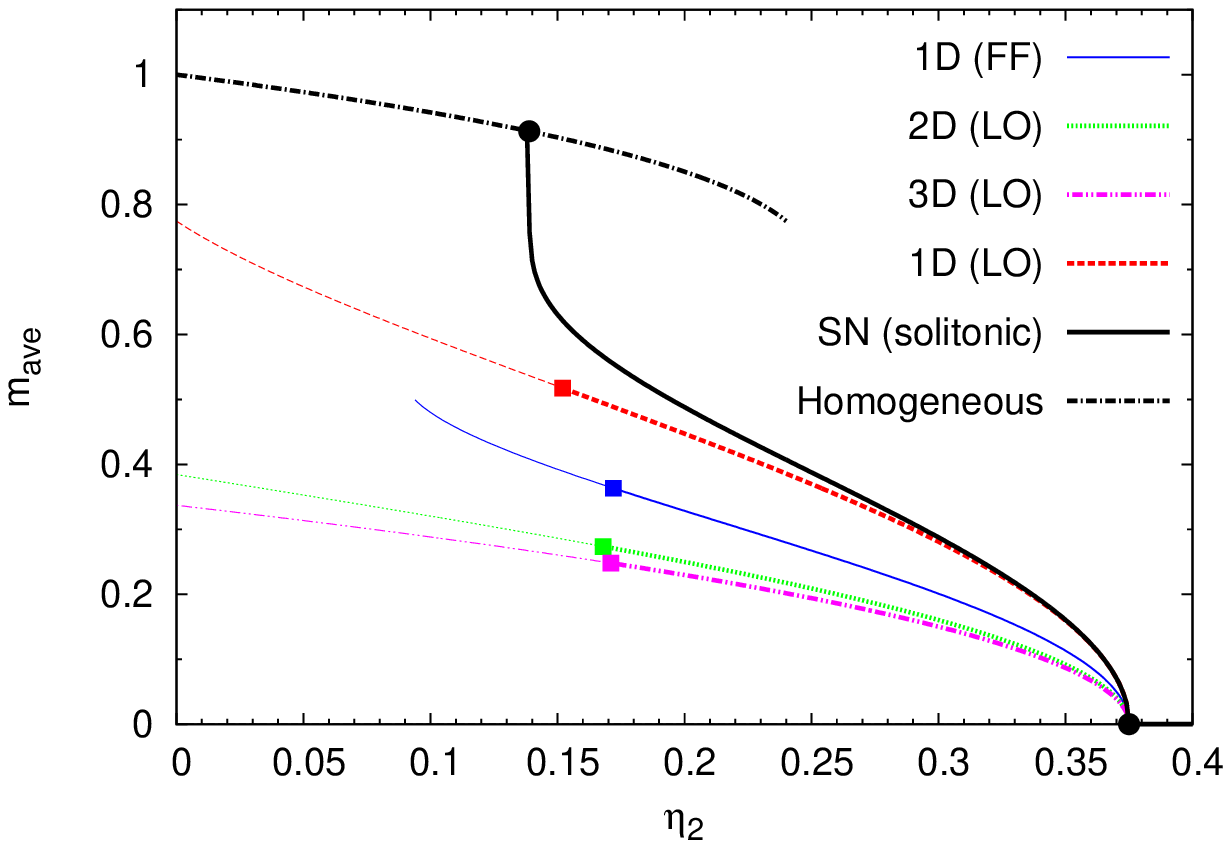}}
\vspace*{-1ex}
\hspace*{-3.15ex}
\subfigure[]{\includegraphics[scale=0.63]{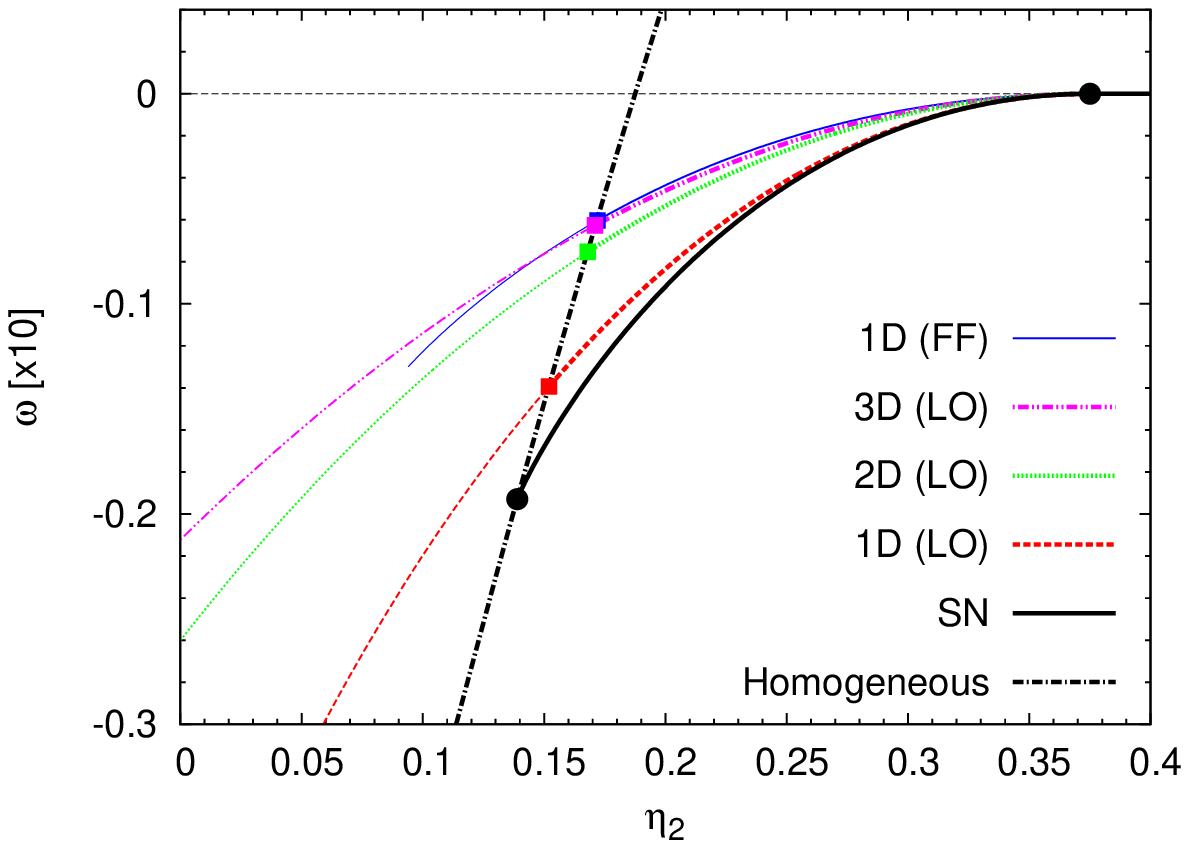}}
\caption{(color online). {\bf (a)}~The effective order parameters for
 the 1D-LO
 (dashed, red), 2D-LO (dotted, green), 3D-LO
 (dot-dot-dashed, magenta), and the solitonic state (solid) as a
 function of  $\eta_2$. 
 For comparison, the quantity for the FF phase is also depicted by
 the thin solid line (blue).
{\bf (b)}~The thermodynamic potentials as a function of $\eta_2$.
}
\label{fig:highdimLO}
\end{figure}
We set multidimensional LO-type real condensates as
\beq
\ba{rcl}
  M_{\mathrm{LO;1D}}({\textbf x})&=&\sqrt{2}M_0\sin(kz),\\[1ex]
  M_{\mathrm{LO;2D}}({\textbf x})&=&M_0(\sin(kx)+\sin(ky)),\\[1ex]
  M_{\mathrm{LO;3D}}({\textbf x})&=&\sqrt{\frac{{2}}{{3}}}M_0%
  (\sin(kx)+\sin(ky)+\sin(kz)).\\[1ex]
\ea
\eeq
$M_{\mathrm{LO;2D}}({\textbf x})$ is equivalent to the ``egg-carton''
{\it Ansatz} in \cite{Carignano:2011gr}.
Each condensate is characterized by two real parameters $M_0$ and $q$
whose values are to be determined via minimization of $\Omega$.
In Fig.~\ref{fig:highdimLO}(a) and (b) we show effective order
parameters and free energies.
One can see that the energy is an increasing function of the
dimensionality. 
Also it is notable that the critical points at which the transitions
from the Wigner to the crystal phases take place are common among all three
states. 
To see this analytically, we expand $\Omega$ in powers of $M_0$ as
\beq
\ba{rcl}
 \Omega_{\mathrm{LO;1D}}&=&\left(\frac{\eta_2}{2}-\frac{3}{16}\right)%
 M_0^2+\frac{6}{24}M_0^4+{\mathcal O}(M_0^6),\\[1ex]
 \Omega_{\mathrm{LO;2D}}&=&\left(\frac{\eta_2}{2}-\frac{3}{16}\right)%
 M_0^2+\frac{9}{24}M_0^4+{\mathcal O}(M_0^6),\\[1ex]
 \Omega_{\mathrm{LO;3D}}&=&\left(\frac{\eta_2}{2}-\frac{3}{16}\right)%
 M_0^2+\frac{10}{24}M_0^4+{\mathcal O}(M_0^6).\\[1ex]
\ea
\eeq
Since the quadratic coefficients take exactly the same form,
$\frac{\eta_2}{2}-\frac{3}{16}\equiv\frac{1}{2}(\eta_2-\eta_2^{\mathrm{II}})$,
the critical point is shared.
On the other hand, from the coefficients of quartic terms, we see that
the free energies are on the order of the dimensionality of modulation.
The above formulas for energy density can be easily generalized to the 
case with the LO condensate in an arbitrary dimension, $d$, defined by
\beq
\textstyle
 M_{\mathrm{LO;dD}}({\textbf x})=\sqrt{\frac{2}{d}}M_0%
  \left(\sin(kx_1)+\cdots+\sin(kx_d)\right),
\eeq
with ${\textbf x}=(x_1,x_2,\cdots,x_d)$ being the $d$-dimensional vector.
In this case the thermodynamic potential looks like
\beq
\ba{rcl}
  \Omega_{\mathrm{LO;dD}}&=&\left(\frac{\eta_2}{2}-\frac{3}{16}\right)%
  M_0^2+\frac{2d-1}{4d}M_0^4\\[1ex]
  &&+\frac{5(4d^2-16d+11)}{48d^2}M_0^6.\\[1ex]
\ea
\eeq
From the dependence of the quartic coefficient on the dimensionality,
$(2d-1)/4d$, we see that the energy is an increasing function of $d$
near $\eta_2=\eta_2^{\mathrm{II}}$.
In fact, after minimizing over $M_0$, the extremal of energy density
becomes
\beq
\ba{rcl}
  \Omega_{\mathrm{LO;dD}}%
  &=&-\left(\frac{\eta_2}{2}-\frac{3}{16}\right)^2\frac{d}{2d-1}%
  +{\mathcal O}\big(\left(\frac{\eta_2}{2}-\frac{3}{16}\right)^3\big).
\ea
\eeq
Note that the higher dimensional LO phases are less favorable than the
1D one, yet they have smaller energies than the 1D-FF (chiral spiral)
state. 

We remark that the ordering of energies with the dimensionality is
not universal, but rather specific to the LO-type. 
To see this fact, we first {\em define} the $d$-dimensional FF-type 
complex condensate as
\beq
 \textstyle
 M_{\mathrm{FF;dD}}({\textbf x})=\sqrt{\frac{1}{d}}M_0%
  \left(e^{ikx_1}+e^{ikx_2}+\cdots+e^{ikx_d}\right).
\eeq
Then after working out similar calculations as in the LO case, we obtain
the potential
\beq
\ba{rcl}
  \Omega_{\mathrm{FF;dD}}&=&\left(\frac{\eta_2}{2}-\frac{3}{16}\right)%
  M_0^2+\frac{1}{2}M_0^4-\frac{4d^2+10d-7}{12d^2}M_0^6.\\[1ex]
\ea
\eeq
In this case we see that the deference first appears at sixth
order. Moreover, the coefficient $(7-10d-4d^2)/(12d^2)$ is no
longer a monotonic function of $d$; it has the absolute minimum at
$d=1.4$.
Above this value, the energy is a monotonically increasing function.
Looking up the values
$(4d^2+10d-7)/(12d^2)\cong\{0.55,0.60,0.58\}$ for $d=\{1,2,3\}$,
we see that the energy density is on the order of
$\Omega_{\mathrm{FF;2D}}<\Omega_{\mathrm{FF;1D}}<\Omega_{\mathrm{FF;3D}}$.

We also tried several other, more exotic {\it Ans\"atze} for the higher
dimensional crystal, including the 2D hexagonal lattices both in the
FF type and the LO type, but could not find the structure with energy
less than the 1D solitonic state. 

\section{Phase structure off the tricritical point}\label{sec:offcri}
We now address the question of how the phase structure changes going away
from the tricritical point.
As long as we restrict ourselves up to sixth order in the GL potential,
the phase structure in the $[\alpha_2,\sgn(\alpha_4)\alpha_4^2]$
plane is simple such that any phase boundary is expressed by a line
starting from the origin.
There is no chance for these lines to meet again away from the origin. 
However this may not be the case in a realistic situation as stated
in the Introduction.
In this section we thus try to extend the previous analysis going beyond
a minimal (sixth order) GL, and see if such nontrivial
phase structure indeed shows up off the tricritical point.

General eighth-order terms can be constructed by differentiating the 
sixth-order terms twice, and collecting terms which are unique up to total
derivatives. 
At order ${\mathcal O}(\nabla^2)$, only $M^4(\nabla M)^2$ comes out, and
at order ${\mathcal O}(\nabla^4)$ three terms [$(\nabla
M)^4,M(\nabla M)^2\Delta M$, and $M^2(\Delta M)^2$] show up. 
Finally at order ${\mathcal O}(\nabla^6)$ only the term $(\nabla\Delta
M)^2$ is possible.
Thus the most general eighth-order terms to be
appended to the GL Lagrangian Eq.~\eqref{eq:GLlag} are collected as
\beq
\ba{rcl}
 \delta\omega_8(M({\textbf x}))&=&\disp\frac{\alpha_8}{8}M^8%
 +\frac{\alpha_{8b}}{8}M^4(\nabla M)^2\\[2ex]
 &&\disp +\frac{\alpha_{8c}}{8}(\nabla M)^4+\frac{\alpha_{8d}}{8}M(\nabla
 M)^2\Delta M\\[2ex]
 &&\disp +\frac{\alpha_{8e}}{8}M^2(\Delta
 M)^2+\frac{\alpha_{8f}}{8}(\nabla\Delta M)^2.\\[1ex]
\ea
\eeq
In NJL-type models, only the quark loops contribute to the GL
coefficients.
The evaluation of them is tough but straightforward to work out. 
At the end of the day we arrive at the linear relations (up to total
derivatives);
\beq
\ba{rcl}
 (\alpha_{8b},\alpha_{8c},\alpha_{8d},\alpha_{8e},\alpha_{8f})%
 =\left(14,-\frac{1}{5},\frac{18}{5},\frac{14}{5},\frac{1}{5}\right)\alpha_8.
\ea
\eeq
Thus it is only $\alpha_8$ that we need to add as a new independent GL
parameter.

Here we briefly repeat the dimensional and scaling analyses in this
case. Since $\alpha_8$ has dimension $\Lambda^{-4}$ and is always
positive, we use it for the energy scale.
Then by use of $|\alpha_6|$ we can again tune the magnitude of the 
sixth- and eighth-order coefficients. To be precise let us introduce the
dimensionless GL couplings $\{\eta_2,\eta_4\}$ and the physical
quantities $\{\tilde{\omega},m,\tilde{\bf x}\}$ via
\beq
\ba{rcl}
  \alpha_2&=&\eta_2\,\left[\frac{|\alpha_6|^3}{\alpha_8^2}\right],\quad%
  \alpha_4=\eta_4\,\left[\frac{\alpha_6^2}{\alpha_8}\right]\\[2ex]
  \omega&=&\tilde{\omega}\,\left[\frac{\alpha_6^4}{\alpha_8^3}\right],\quad %
  M=m\,\left[\frac{|\alpha_6|^{1/2}}{\alpha_8^{1/2}}\right],\\[2ex]
  {\bf x}&=&\tilde{\bf
  x}\,\left[\frac{\alpha_8^{1/2}}{|\alpha_6|^{1/2}}\right],%
  \quad \tilde{\nabla}=\nabla_{\tilde{\bf x}}.\\[1ex]
\ea
\eeq
Then in these dimensionless bases, we have
\beq
\ba{rcl}
 \tilde{\omega}&=&\frac{1}{2}\eta_2m^2%
 +\frac{1}{4}\eta_4\Big(m^4+(\tilde{\nabla}m)^2\Big)\\[1ex]
  &&+\frac{\sgn(\alpha_6)}{6}\Big(%
  m^6+5m^2(\tilde{\nabla}m)^2+\frac{1}{2}%
  (\tilde{\Delta}m)^2\Big)\\[1ex]
  &&+\frac{1}{8}\Big(m^8+14m^4(\tilde{\nabla}m)^2-\frac{1}{5}(\tilde\nabla
  m)^4\\[1ex]%
  &&+\frac{18}{5}m(\tilde\nabla m)^2\tilde\Delta m%
  +\frac{14}{5}\tilde m^2(\tilde{\Delta}m)^2%
  +\frac{1}{5}(\tilde\nabla\tilde\Delta m)^2\Big).
\ea
\eeq
In this case, we have two independent dimensionless couplings
$(\eta_2,\eta_4)$.
When $\alpha_6>0$ the limit $\alpha_8\to 0^+$ can be taken, and in this
limit the thermodynamics should not depend on $\alpha_8$ so that the
physical quantities depend on $(\eta_2,\eta_4)$ only through the
combination of $\eta_2/\eta_4^2=\alpha_2\alpha_6/\alpha_4^2$.
This is the realization of $(\alpha_2/\alpha_4^2)$ scaling
in the absence of eighth-order terms that results in a linear relation
between $\alpha_2$ and $\alpha_4^2$ at any phase transition.
This translates into a linear relation between $\eta_2$ and $\eta_4^2$
in the limit of vanishing $\alpha_8$.
We now expect a nonlinear relation between $\eta_4^2$ and $\eta_2$ due to
the effect of $\alpha_8$ by going away from the origin.
From now on we throw away tildes on physical quantities, so we should keep
in mind that they are dimensionless and properly rescaled quantities.

\subsection{Phase diagram for $\alpha_6>0$}
\begin{figure}[t]
\centering
\includegraphics[scale=0.65]{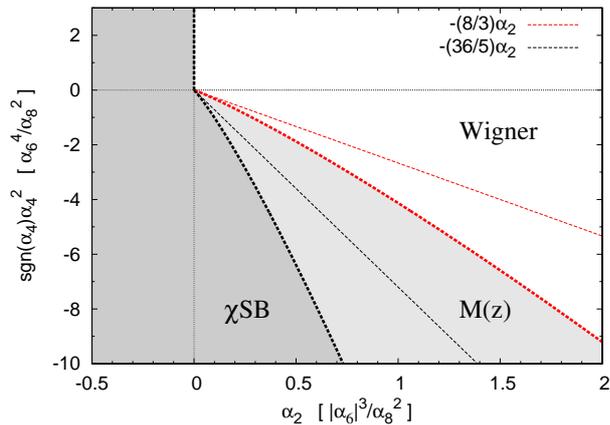}
\caption{(color online). Phase diagram for $\alpha_6>0$:~The phase
 boundaries in the
 absence of eighth-order terms (see Fig.~\ref{fig:phasediagram}) are
 also repeated by thin dashed lines, for comparison. 
}
\label{fig:phasediagram8}
\end{figure}

We begin the discussion with the case of $\alpha_6>0$ which has a close
relation with the previous analyses with $\alpha_8=0$.
The phase diagram for this case is displayed in
Fig.~\ref{fig:phasediagram8}. 
The phase boundaries are depicted by dashed lines. 
All the transitions are of second order.
Just for comparison, we also show the phase boundaries obtained 
for the case $\alpha_8=0$ with two thin dashed lines, which correspond
to two lines in Fig.~\ref{fig:phasediagram}.
As expected, we see that two critical lines are no longer linear in the
plane, which is a simple consequence of the violation of the
$(\eta_2/\eta_4^2)$ scaling. 
We can derive the analytic formulas for two critical lines.
Analyzing the condition of the soliton formation
$\min_{\{M_0,k\}} f(M_0,k,\eta_2,\eta_4)\to 0^+$ with $f$ defined by
\beq
  f(M_0,k,\eta_2,\eta_4)\equiv%
  \frac{d\langle\tilde{\omega}(M_{\mathrm{sn}}(z))%
  \rangle}{d\nu}\Big|_{\nu\to 1^-},
\eeq
we easily arrive at the analytical expression for the soliton onset
in a closed compact form, which can be expanded in powers of $\eta_4<0$
as 
\beq
  \eta_2=\frac{5}{36}\eta_4^2+\frac{125}{1512}\eta_4^3%
  +\frac{3125}{28224}\eta_4^4+\cdots.
\eeq
By use of original couplings, this translates into
\beq
  \alpha_2\alpha_6=\frac{5}{36}\alpha_4^2%
  +\frac{125\alpha_8}{1512\alpha_6^2}%
  \alpha_4^3+\frac{3125\alpha_8^2}{28224\alpha_6^4}%
  \alpha_4^4+\cdots,
\eeq
from which we clearly see that the eighth-order coefficient $\alpha_8$ is
responsible for the nonlinearity between $\alpha_2$ and $\alpha_4^2$.
In the same way, we can also derive the compact formula for the critical
line dividing the Wigner and inhomogeneous phases. This can be expanded
for $\eta_4<0$ as
\beq
 \eta_2=\frac{3}{8}\eta_4^2+\frac{27}{160}\eta_4^3%
 +\frac{2187}{12800}\eta_4^4+\cdots.
\eeq
Again the terms equal to or higher than the cubic term are caused by
nonvanishing $\alpha_8$.
We found that, despite these modifications of two critical lines, the
qualitative feature of the phase diagram remains unchanged. In
particular, the inhomogeneous phase is still dominated by the
one-dimensional modulation characterized by the elliptic function
$M_{\mathrm{sn}}(z)=M_0\nu\mathrm{sn}(kz,\nu)$ with $k=M_0$.

\subsection{Going to new regime: Phase diagram for $\alpha_6<0$}
\begin{figure}[t]
\centering
\includegraphics[scale=0.65]{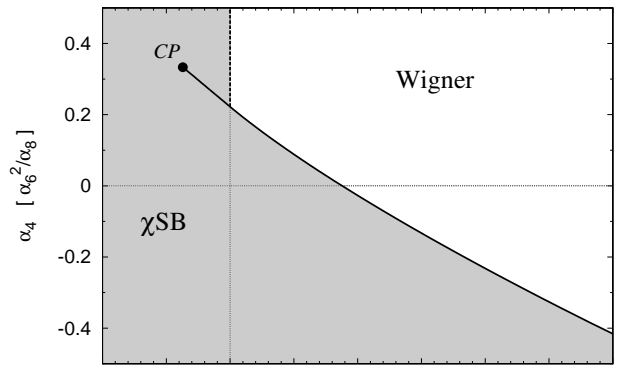}
\includegraphics[scale=0.65]{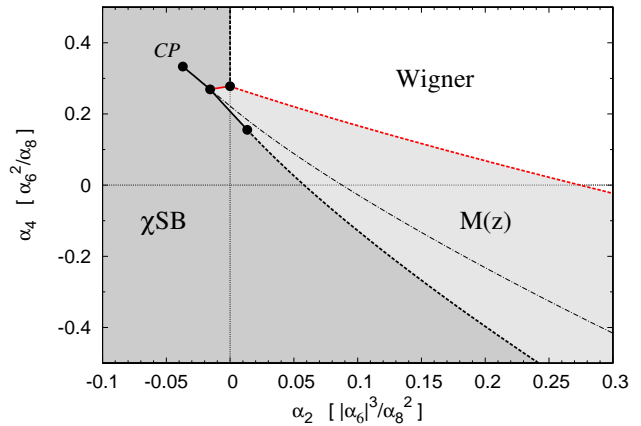}
\caption{Phase diagrams for $\alpha_6<0$: 
The horizontal (vertical) axis is $\alpha_2$ $(\alpha_4)$ in the
 appropriate units shown in the parentheses, which is actually
 equivalent to the dimensionless coupling, $\eta_2$ ($\eta_4$).
The phase diagram obtained
 with restriction to the homogeneous condensate (upper panel) and that
 computed without limitation to the homogeneous condensate (lower panel).
 Solid lines refer to first-order phase transitions, while dashed lines
 represent the second-order phase transitions.
}
\label{fig:phasediagram8m1}
\end{figure}

We now have a closer look at the case $\alpha_6<0$ which is not
connected in any limits of the previous analysis with $\alpha_8=0$.
In Fig.~\ref{fig:phasediagram8m1}, we show the phase diagrams for
$\alpha_6<0$ with and without restriction to the homogeneous condensate. 

The upper panel shows the phase diagram computed with restriction to the
homogeneous condensate.
In this case, we have first-order chiral restoration even for $\alpha_4>0$
because of negative $\alpha_6$.
Moreover when $\alpha_4>9/2$ there are two types of $\chi$SB phases,
one with a larger chiral condensate, and the other with a smaller condensate
which we label as the ``$\chi$SB${}_2$'' phase.
The phase transition from the $\chi$SB${}_2$ phase to the Wigner phase
is second order. On the other hand, the transition between the $\chi$SB and
the $\chi$SB${}_2$ phases is first order.
The magnitude of the chiral condensate jumps across the first-order phase
transition. 
The gap between two condensates at the transition decreases with
increasing $\alpha_4$, and it eventually ends at the critical point
marked by \emph{CP}.
The coordinate of this critical point is found as $(\eta_2,\eta_4)%
=(-1/27,1/3)$. Accordingly, \emph{CP} scales with $\alpha_6$ as
\beq
\ba{rcl}
\mathrm{\emph{CP}}&:&(\alpha_2,\alpha_4)%
\ea
=\left(%
\disp-\frac{1}{27}\frac{|a_6|^3}{a_8^2},\, %
\disp\frac{1}{3}\frac{a_6^2}{a_8}\right),
\eeq
from which it is clear that point \emph{CP} goes to the origin
$(0,0)$ when $\alpha_6\to 0^-$; it smoothly continues to the Lifshitz
point in the region $\alpha_6>0$.
At point \emph{CP}, the thermodynamic potential takes the following form in
the dimensionless units:
\beq
 \tilde{\omega}=-\frac{1}{648}+\frac{1}{8}\left(m^2-\frac{1}{3}\right)^4.
\eeq
We see that the sixth-order derivative of $\tilde{\omega}$ with respect to
the order parameter $m$ vanishes at this point. We expect
large thermal fluctuation effects in the vicinity of \emph{CP}.

The lower panel of the figure shows the phase diagram where the
limitation to the homogeneous condensate is taken away. 
A major part (but not all) of the first-order phase transition splits into
two transition lines, and an inhomogeneous phase shows up in between
which is dominated again by one-dimensional modulation characterized by
the elliptic function.
Point \emph{CP} remains unaffected by the inclusion of an inhomogeneous
condensate.

\begin{figure}[t]
\centering
\includegraphics[scale=0.65]{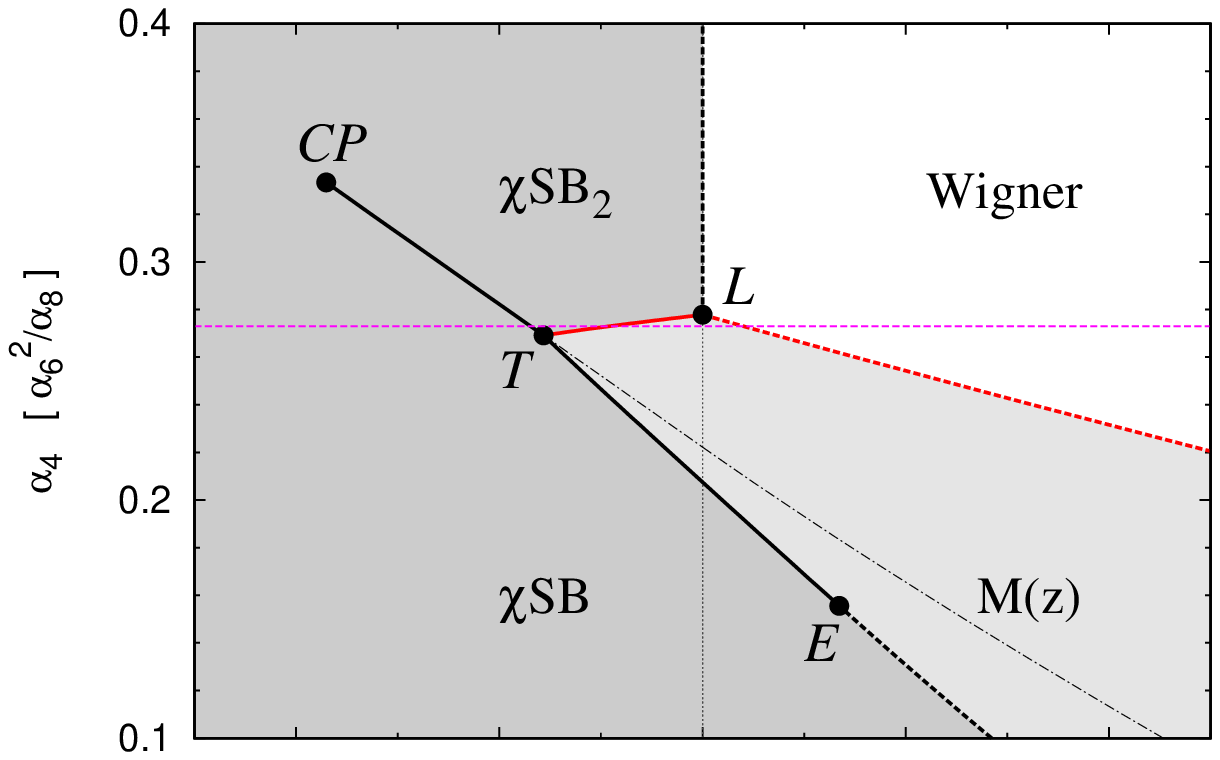}
\includegraphics[scale=0.65]{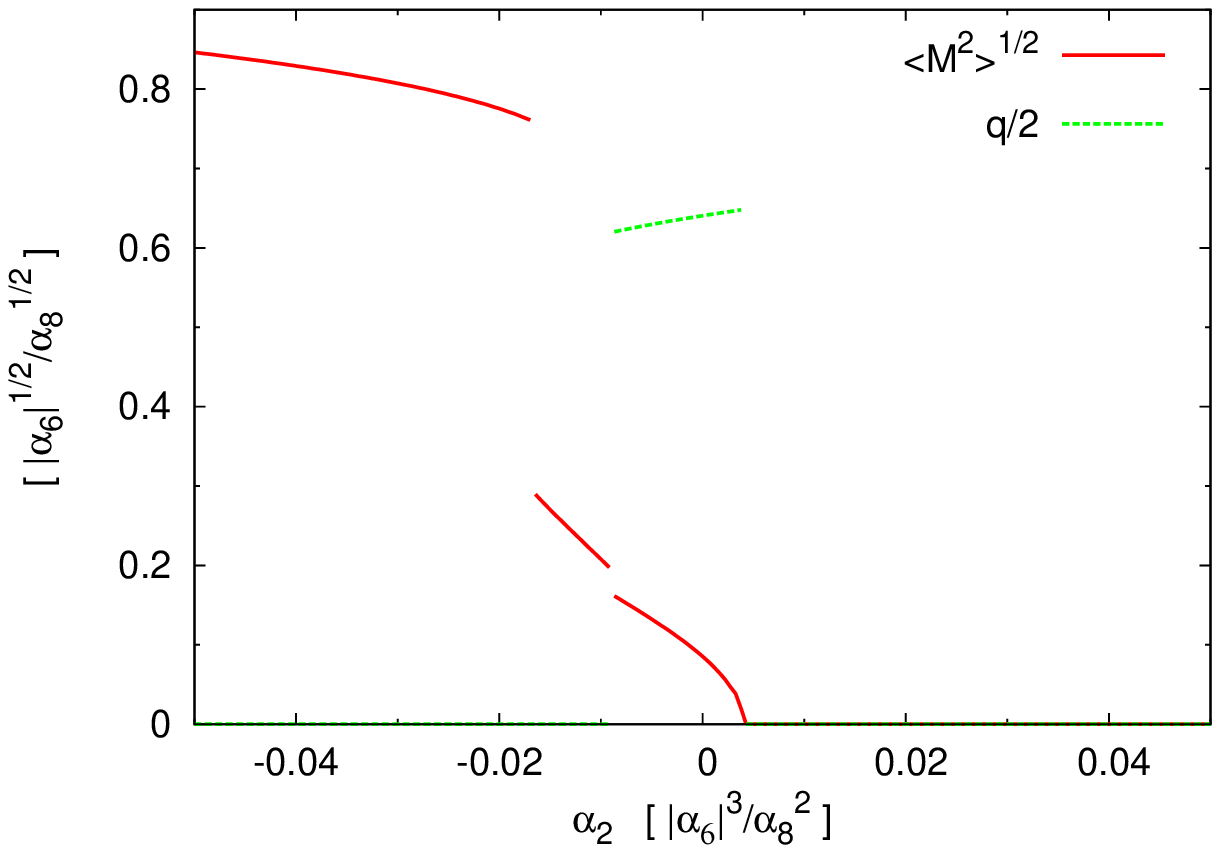}
\caption{(color online). (Upper panel)~A magnified version of the phase diagram in
the $(\alpha_2,\alpha_4)$ plane for $\alpha_6<0$ in the lower panel of
 Fig.~\ref{fig:phasediagram8m1}.
(Lower panel)~The order parameter $\sqrt{\langle M(z)^2\rangle}$, and
 $q=2\pi/\ell_{\mathrm{P}}$ as a function of
 $\alpha_2$ along the line $\alpha_4=0.273$ which is displayed in the
 upper phase diagram by the dashed (magenta) line.
}
\label{fig:phasediagram8m2}
\end{figure}

In addition to \emph{CP}, three points show up in the phase diagram.
Let us closely investigate these points in the following.
In the upper panel of Fig.~\ref{fig:phasediagram8m2}, we present the
magnified figure of the phase diagram.
First, we notice that the Lifshitz point (\emph{L}) moved to
$\alpha_4>0$ from the origin.
The situation is the same at the point in which three phases, the
homogeneous broken phase, the inhomogeneous phase, and the Wigner phase,
meet up.
However, it now has the branch of the first-order transition separating
the inhomogeneous phase and the homogeneous $\chi$SB${}_2$ phase.
The exact location of point \emph{L} can be found analytically as
$(\eta_2,\eta_4)=(0,5/18)$ that means 
\beq
\ba{rcl}
\mathrm{\emph{L}}&:&(\alpha_2,\alpha_4)%
=\left(%
\disp 0,\, %
\disp\frac{5}{18}\frac{a_6^2}{a_8}\right).
\ea
\eeq
Point (\emph{L}) also will become degenerate with the origin as
$\alpha_6\to 0^-$; it extends to the Lifshitz point in the region
$\alpha_6>0$.

There is also the critical end point (\emph{E}) on the critical line dividing
the inhomogeneous phase and the homogeneous broken phase.
At this point the second-order phase transition turns into the 
first-order one. The location of this point can be numerically found as
$(\eta_2,\eta_4)=(0.014,0.16)$, that is,
\beq
\ba{rcl}
\mathrm{\emph{E}}&:&(\alpha_2,\alpha_4)%
=\left(%
\disp 0.014\frac{|a_6|^3}{a_8^2},\, %
\disp 0.16\frac{a_6^2}{a_8}\right).
\ea
\eeq

The most intriguing point that appears for $\alpha_6<0$ is the triple
point denoted by \emph{T} in the figure. At this point, three first-order
phase transitions meet at once. 
The location is found numerically as $(\eta_2,\eta_4)=(-0.016,0.27)$
which corresponds to
\beq
\ba{rcl}
\mathrm{\emph{T}}&:&(\alpha_2,\alpha_4)%
=\left(%
\disp -0.016\frac{|a_6|^3}{a_8^2},\, %
\disp 0.27\frac{a_6^2}{a_8}\right).
\ea
\eeq
Three different forms of the chiral phase compete with each other and coexist at
this point. Any dimensionless ratio approaches to some constant when
point \emph{T} is approached.
We found numerically the following values
\beq
\ba{rcl}
\disp\lim_{(\eta_2,\eta_4)\to\mathrm{\emph{T}}}\frac{M(\chi\mathrm{SB}_2)}{M(\chi\mathrm{SB})}&=&0.369\cdots,\\[2ex]
\disp\disp\lim_{(\eta_2,\eta_4)\to\mathrm{\emph{T}}}\frac{\sqrt{\langle
M(z)^2\rangle}}{M(\chi\mathrm{SB})}&=&0.308\cdots,\\[2ex]
\disp\disp\lim_{(\eta_2,\eta_4)\to\mathrm{\emph{T}}}\frac{q}{\sqrt{\langle M(z)^2\rangle}}&=&5.001\cdots,\\[1ex]
\ea
\eeq
where $q=2\pi/\ell_{\mathrm{P}}$ with $\ell_{\mathrm{P}}$ being the
elliptic period.
All these ratios are universal in the sense that they do not depend on
any of the GL couplings.
Thus we would say these ratios are associated with the triple point
(\emph{T}) itself. In other words, these values characterize the point
(\emph{T}). Therefore the values obtained here are model independent and should
be shared by all theories which allow the existence of such triple point.

All four points, \emph{CP}, \emph{T}, \emph{L}, and \emph{E}, discussed above get shrunk
to the origin when the limit $\alpha_6\to 0^-$ is taken, and they
continue to the Lifshitz point in the region $\alpha_6>0$.

In the lower panel of Fig.~\ref{fig:phasediagram8m2}, we show
the effective order parameter $\sqrt{\langle M(z)^2\rangle}$, and $q$ as
a function of $\alpha_2$ along the section $\eta_4=0.273$ which is
depicted by a horizontal dashed (magenta) line in the phase diagram
displayed in the upper panel.
We see that the chiral restoration in this case proceeds via three steps:
first from the $\chi$SB to the $\chi$SB${}_2$ phase, second from the
$\chi$SB${}_2$ to the inhomogeneous phase, and last from the
inhomogeneous phase to the Wigner phase through second-order phase
transition.

\subsection{How does the GL map onto the $(\mu,T)$ phase diagram?}
\begin{figure}[t]
\centering
\includegraphics[scale=0.65]{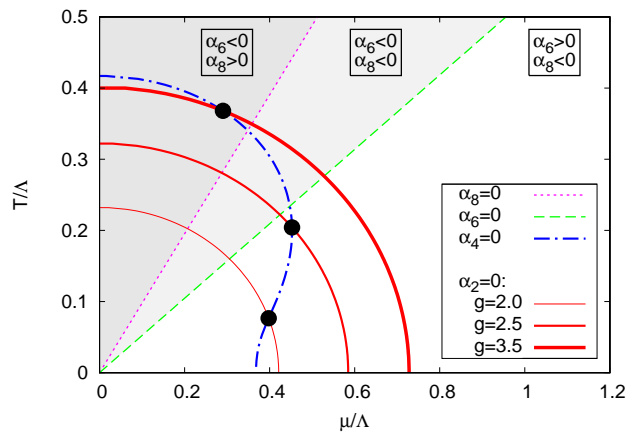}
\caption{(color online). Mapping of the conditions of vanishing GL
 coefficients to the $(\mu,T)$-phase space. 
 Solid (red) lines correspond to $\alpha_2=0$ for several
 different values of the NJL coupling; $g=G\Lambda^2$, $2.0$ (light), $2.5$
 (medium), and $3.5$ (heavy). The dot-dashed (blue) line presents
 the condition $\alpha_4=0$.
 The intersection of the lines, $\alpha_2=0$ and $\alpha_4=0$,
 provides the candidate of the critical point. 
 Area shadings are made according to the signs of $\alpha_6$ and
 $\alpha_8$.
}
\label{fig:mapping}
\end{figure}
The microscopic evaluation of GL couplings within the NJL model can be
summarized in the form \cite{Nickel:2009ke}
\begin{equation}
\textstyle
 \alpha_{2n}%
 =\frac{\delta_{n,1}}{2G}+4N_cN_fT\sum_m\int\frac{d\bm{p}}{(2\pi)^3}%
 \frac{1}{((i\omega_m+\mu)^2-p^2)^n},
\label{eq:sum}
\end{equation}
where $\omega_m=\pi T(2m+1)$ is the fermionic Matsubara frequency. The
summation over the frequency can be done, resulting in the following
expressions for $\alpha_2$ and $\alpha_4$
\begin{subequations}
\begin{eqnarray}
 \alpha_2&=&\textstyle\frac{1}{2G}-4N_cN_f\int\frac{d\bm{p}}{(2\pi)^3}%
 \frac{1-f_F(p-\mu)-f_F(p+\mu)}{2p},\\[2ex]
 \alpha_4&=&\textstyle N_cN_f\int\frac{d\bm{p}}{(2\pi)^3}%
 \frac{1-f_F(p-\mu)-f_F(p+\mu)}{p^3}\\[2ex]
 &&\textstyle%
 +N_cN_f\int\frac{d\bm{p}}{(2\pi)^3}\frac{f_F'(p-\mu)+f_F'(p+\mu)}{p^2}.%
 \notag
\end{eqnarray}
\end{subequations}
We can derive similar expressions for higher order coefficients.
The momentum integral in $\alpha_2$ $(\alpha_4)$ has quadratic
(logarithmic) divergence so that we need a cutoff $\Lambda$.
As a consequence, their functional forms become
$\alpha_2=\Lambda^2f_1(\mu/\Lambda,T/\Lambda,g)$ and
$\alpha_4=f_2(\mu/\Lambda,T/\Lambda)$
with $g\equiv G\Lambda^2$ as the dimensionless NJL coupling
and $\{f_n\}$ being some dimensionless functions.
Higher order coefficients, $\alpha_{2n}$ $(n\ge 3)$,
no longer have the UV divergence so that their functional forms can be
summarized as $\alpha_{2n}(\mu,T)=f_n(\mu/T)/\mu^{2(n-2)}$.
$f_1$ ($f_2$) is a function having three (two) arguments, while $f_{n}$
for $(n\ge3)$ has only a single argument, $\mu/T$.
Thus the condition of vanishing $\alpha_{2n}$ for $n\ge 3$ turns into
a straight line in the $(\mu,T)$ plane if $f_{n}(x)=0$ has a solution
for $x\ge 0$.

\begin{figure}[tp]
\centering
\includegraphics[scale=0.463]{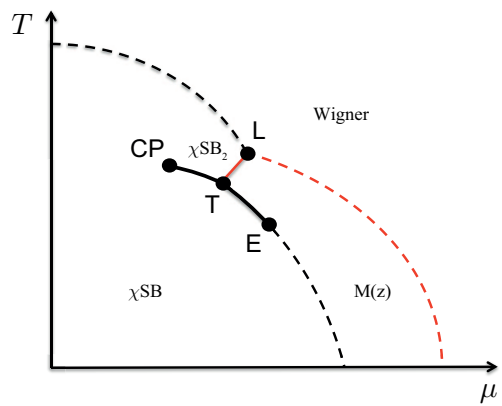}
\caption{Conjectured, schematic phase diagram. 
 The Lifshitz point splits into four critical points: the Lifshitz point,
 the critical point, the end point, and the triple point, labeled by
 \emph{L}, \emph{CP}, \emph{E}, and \emph{T}, respectively.
}
\label{fig:phaseconjecture}
\end{figure}

It is easy to evaluate the vanishing conditions of $\alpha_{2,4,6,8}$,
and the results are depicted in the $(\mu/\Lambda,T/\Lambda)$ plane in
Fig.~\ref{fig:mapping}.
Solid (red) curves show the vanishing of $\alpha_2$ for
$g=2.0$ (light), $g=2.5$ (medium), and $g=3.5$ (heavy), respectively.
The dot-dashed (blue) curve represents the vanishing of
$\alpha_4$. 
The intersection of these two curves, where $\alpha_2=\alpha_4=0$,
presents the candidate of the location of the critical point,
in the absence of inhomogeneous condensates.
We see that the location shifts to higher temperatures with increasing
the strength of the NJL coupling, $g$, while its $\mu$ coordinate shows
nonmonotonic dependence on $g$.

The long-dashed (green) line and the dashed (magenta) line
show the lines for vanishing $\alpha_6$ and $\alpha_8$, respectively.
Accordingly, the condition ($\alpha_8>0$, $\alpha_6<0$) is realized in
the shaded area, while $\alpha_8$ changes its sign in the lightly shaded
area where ($\alpha_8<0$, $\alpha_6<0$) is satisfied.
The region without shading corresponds to ($\alpha_6>0$, $\alpha_8<0$).
The present analysis suggests that if the ``would-be'' critical point is
located in the shaded area, this point may split to the four independent
points, \emph{CP}, \emph{T}, \emph{L}, and \emph{E} discussed in the
previous section. 
If this is the case, the corresponding $(\mu,T)$-phase diagram might become
the one schematically depicted in Fig.~\ref{fig:phaseconjecture}

We remark that this interesting situation is realized for very large
values of the NJL coupling.
The NJL coupling empirically has taken ranges from $g\cong 1.8$ to
$g\cong2.5$ depending on the presence/absence of the 't Hooft coupling
\cite{Hatsuda:1994pi,Klevansky:1992qe,Lutz:1992dv,Buballa:2003qv}.
Then it would be unlikely to have the triple point in the standard
version of the NJL model.
It would be interesting to investigate if the situation could be
realized in the extended version of the NJL models such as one
incorporated with the Polyakov-loop
\cite{Fukushima:2003fw,Ratti:2005jh,Abuki:2008tx,%
Abuki:2008ht,*Abuki:2008nm,Abuki:2009dt}, 
the other type interactions, the diquark degrees of freedom, etc.

\section{Summary}\label{sec:summary}
We have made an extensive analysis on inhomogeneous chiral
condensates using the Ginzburg-Landau approach.
We first reviewed the work by Nickel \cite{Nickel:2009ke}
with special emphasis put on the mechanism of the formation of a 
chiral domain wall.
We have also discussed singular behaviors of the thermodynamic
quantities in the vicinity of soliton formation.

We then analyzed the most general condensate of the 1D modulation,
$M_{\mathrm{HH}}(z)$ expanded in harmonics.
We confirmed that $M_{\mathrm{HH}}$ seems to converge to the
solitonic state characterized by Jacobi's elliptic function
by increasing the number of harmonics.
It well approximates the solitonic condensate except for in the vicinity
of a domain wall onset, already at the level of truncation with the
first few harmonics. 
This strongly suggests that the elliptical solitonic state is the most
favorable structure within 1D modulations.

We then explored the possibility of realization of multidimensional chiral
crystals. 
Taking several {\it Ans\"atze} for multidimensional structures, we
compared free energies. 
We have shown that, in the case of real condensates of the LO-type, the
free energy density is an increasing function of the dimensionality of
modulation $d$.
On the other hand, it has been confirmed that the energy takes a
minimum for $d=2$
in the case of complex condensate of the FF-type.
We derived analytic expressions for the effective potentials, both
for the real and complex condensates forming a simple square and cubic
lattices.
The order of energy was correctly understood based on the GL
expansion near the critical point.

Since we have restricted our analyses to simple lattice structures,
the quest for such crystal structures remains the subject of
continuing interest.
In particular, because simple 1D structures are known to be unstable
against the thermal fluctuation \cite{Baym:1982ca}, we should clarify
how the 1D structure could be stabilized if we really have no suitable
multidimensional structure.

We have investigated the phase structure away from the tricritical
point by extending the previous GL functional to one expanded
on the order parameter and its spatial derivative up to eighth order.
With this extension, we analyzed phase structures both for
$\alpha_6>0$ and $\alpha_6<0$. 
For $\alpha_6>0$, we have seen that two critical lines enclosing the
inhomogeneous phase change their slopes in the
$[\alpha_2,\sgn(\alpha_4)\alpha_4^2]$ plane by going away
from the Lifshitz point.
This is a natural consequence of including the eighth-order terms.
We have derived the analytical expressions for the two critical curves.
However no new phase structure was found to appear even 
largely departing from the Lifshitz point.

On the other hand, once $\alpha_6$ changes its sign to negative,
$\alpha_6<0$, several qualitative changes appear immediately in the
$(\alpha_2,\alpha_4)$-phase diagram.
We have seen that the Lifshitz point located at the origin for
$\alpha_6>0$ splits into four critical points; these are the critical
point (\emph{CP}), the triple point (\emph{T}), the end point (\emph{E}),
and the new Lifshitz point (\emph{L}), which has a branch of the
first-order phase transition.
The distances among these points grow with increasing the magnitude
of $|\alpha_6|$.
There are two kinds of homogeneously symmetry-broken phases in this case:
one with a larger chiral condensate $\chi$SB, and the other
with a smaller chiral condensate which we label by $\chi$SB${}_2$.
They are separated by a first-order phase transition extending from the
triple point (T) to the critical point (CP) at which the phase
transition terminates.
Most intriguing is the emergence of the triple point (T) where three
different forms of the chiral phase, $\chi$SB, $\chi$SB${}_2$ and the
inhomogeneous (solitonic) phase, are present.
Any dimensionless ratios of physical quantities approach some
specific values when the triple point (T) is approached.
These values are universal, model independent, being associated with the
triple point itself.
We have extracted numerically the universal ratios associated with the
triple point (T).

There are several possible extensions of the current work.
We have restricted our analyses to the chiral limit. It is
straightforward to extend our work so as to include the effect
of a finite current quark mass along the line of \cite{Boehmer:2007ea}. 
It may also be interesting to try the most general complex condensate,
the twisted chiral crystal \cite{Basar:2008im,Basar:2008ki},
even though a simple chiral spiral condensate of
the FF type turned out to be the most favorable structure within a
1D-NJL model \cite{Basar:2009fg}.
Also the interplay between the chiral condensate and color
superconductivity is of particular interest since it was shown
that there is an entanglement between them via the axial anomaly in QCD
\cite{Hatsuda:2006ps}.
Its effects have been proven to be so drastic that it brings a rich
phase structure near the critical point 
\cite{Hatsuda:2006ps,Yamamoto:2007jn,Abuki:2010jq,Basler:2010xy}.
A more interesting, but challenging issue is the first principle
determination of the Ginzburg-Landau coefficients with the lattice QCD.
Extension of the present work in these directions deserves future work.

\begin{acknowledgments}
T.~Brauner, Y.~Hidaka, Y.~Iwata, K.~Kamikado, T.~Kojo
 A.~Ohnishi, D.~Rischke, M.~Ruggieri and J.~Wambach are acknowledged
 for their interests in this work and several suggestive comments.
We thank all the members of the theoretical quark/hadron group at
 Tokyo University of Science for useful conversations. 
\end{acknowledgments}

\bibliographystyle{apsrev4-1}
\bibliography{refsmin}

\end{document}